\documentclass[%
 reprint,
 amsmath,amssymb,
 aps,
]{revtex4-2}

\usepackage{graphicx}
\usepackage{dcolumn}
\usepackage{bm}
\usepackage{xcolor}
\usepackage{comment}

\begin{document}

\preprint{APS/123-QED}

\title{Design of a multi-channel photonic crystal dielectric laser accelerator}

\author{Zhexin Zhao}
\author{Dylan S. Black}
\author{Tyler W. Hughes}
\author{Yu Miao}
\author{Olav Solgaard}
\author{Robert L. Byer}
\author{Shanhui Fan}
 \email{shanhui@stanford.edu}
\affiliation{Ginzton Laboratory, 348 Via Pueblo, Stanford University, Stanford, CA 94305}
\author{R. Joel England}
\affiliation{SLAC National Accelerator Laboratory, 2575 Sand Hill Road, Menlo Park, CA 94025}
\collaboration{ACHIP Collaboration}

\date{\today}

\begin{abstract}
To be useful for most scientific and medical applications, compact particle accelerators will require much higher average current than enabled by current architectures. For this purpose, we propose a photonic crystal architecture for a dielectric laser accelerator, referred to as a multi-input multi-output silicon accelerator (MIMOSA),
that enables simultaneous acceleration of multiple electron beams, increasing the total electron throughput by at least one order of magnitude. To achieve this, we show that the photonic crystal must support a mode at the $\Gamma$ point in reciprocal space, with a normalized frequency equal to the normalized speed of the phase matched electron. We show that the figure of merit of the MIMOSA can be inferred from the eigenmodes of the corresponding infinitely periodic structure, which provides a powerful approach to design such devices.
Additionally, we extend the MIMOSA 
architecture to electron deflectors and other electron manipulation functionalities. These additional functionalities, combined with the increased electron throughput of these devices, permit all-optical on-chip manipulation of electron beams in a fully integrated architecture compatible with current fabrication technologies, which opens the way to unconventional electron beam shaping, imaging, and radiation generation.
 
\end{abstract}

\maketitle

\section{Introduction}
Dielectric laser accelerators (DLAs) accelerate charged particles using the evanescent fields of dielectric structures driven by femtosecond laser pulses
\cite{plettner2006proposed, england2014dielectric, peralta2013demonstration, breuer2013laser, breuer2014dielectric}. 
Due to the high damage threshold of dielectric materials in the near-infrared, the acceleration gradient of DLAs can be more than ten times higher than conventional radio-frequency accelerators \cite{cesar2018enhanced, cesar2018high, wootton2016demonstration, leedle2015dielectric}. Recent progress in the integration of DLAs with photonic circuits has enabled the development of chip-scale accelerators, 
\cite{hughes2017method, hughes2017chip, hughes2019reconfigurable, tan2019silicon, zhao2018design, sapra2019chip}, which promise to further increase the compactness and practicality of this scheme. 
\textcolor{black}{For sub-relativistic and moderately relativistic electrons, focusing provided by laser-driven techniques, such as alternating phase focusing \cite{niedermayer2018alternating, naranjo2012stable}, enables stable electron beam transport and acceleration for long distance.}

\begin{figure}
    \centering
    \includegraphics[width=0.45\textwidth]{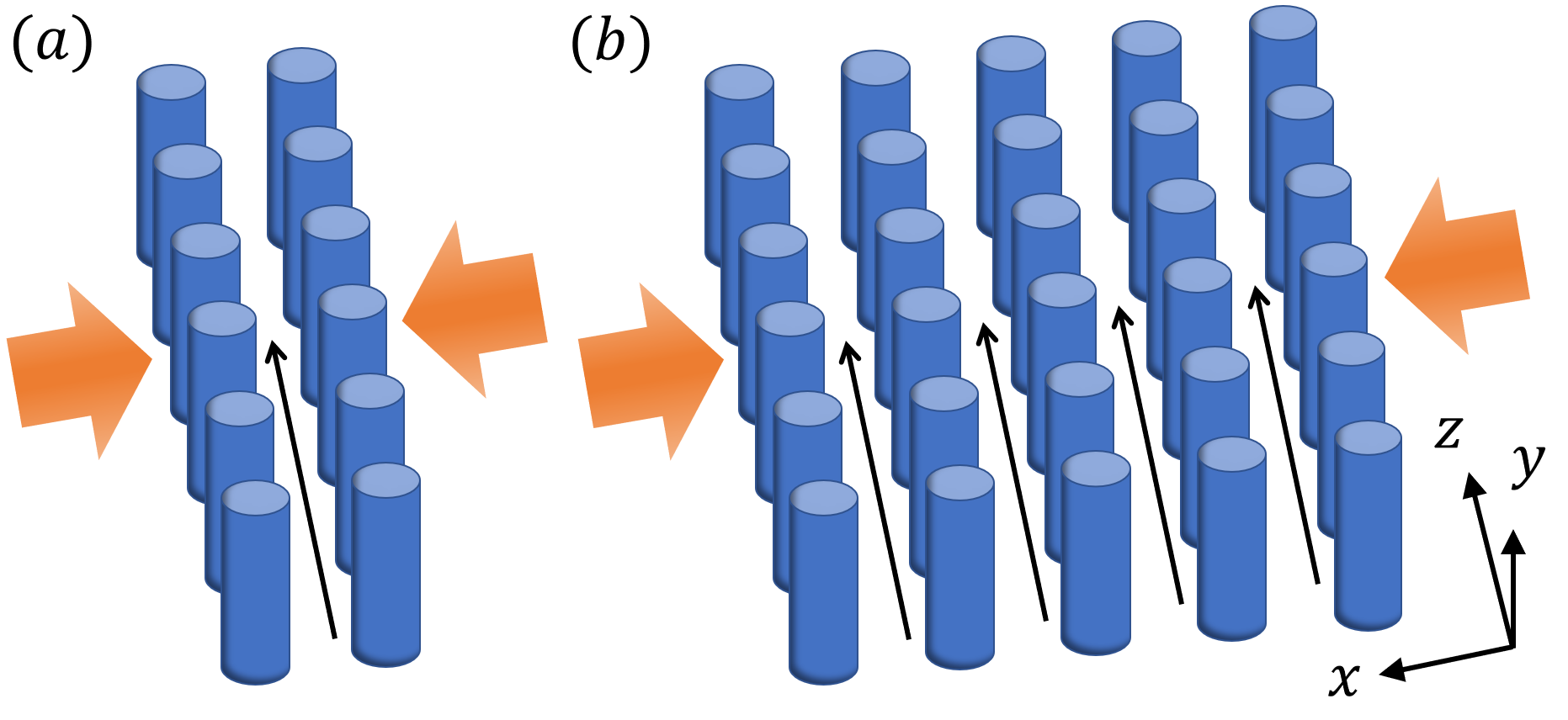}
    \caption{Schematic of a dual pillar DLA (a) and a multi-channel DLA (b). Two laser pulses (propagating in $\pm x$ directions) incident on the DLA are indicated by the orange arrows. Electrons travel inside parallel channels along \textit{z} direction.}
    \label{figschematic}
\end{figure}

To promote the application of DLAs in both fundamental science and medical therapy \cite{england2014dielectric, wootton2016dielectric, ody2017flat}, it is important to deliver high electron currents. However, since the geometric dimensions of currently existing DLA designs are on the wavelength scale, where the wavelength corresponds to that of their near-infrared drive lasers, it is intrinsically challenging to deliver an electron beam with high current through a single narrow electron channel as is currently used in DLAs \cite{niedermayer2018alternating}. 
Motivated by this challenge, we explore a photonic crystal DLA architecture that has multiple electron channels (Fig. \ref{figschematic}). We show that straightforward design principles lead to the electromagnetic fields inside different channels being almost identical, which enables simultaneous acceleration or manipulation of N phase-locked electron beams, increasing the total current by a factor of N.

\section{Design principles}
Typical DLAs consist of a pair of dielectric gratings, such as the dual silicon pillar DLAs illustrated in Fig. \ref{figschematic}(a) \cite{leedle2015dielectric}. When the laser beams illuminate the gratings, the near fields can be used to accelerate the electron beam that travels in the gap between two gratings. Although the height of the electron channel can be several micrometers, its width is limited to the sub-micrometer scale, due to the evanescent nature of the near fields. The small channel width limits the total electron current. To bypass this constraint, we propose a DLA architecture that consists of multiple parallel electron channels, as shown in Fig. \ref{figschematic}(b). \textcolor{black}{The idea of simultaneous acceleration of multiple parallel beams has also been studied in millimeter wave accelerators \cite{whittum2001switched, whittum1998active, zimmermann1998new}. However, the underlying physics and design principles of the multi-channel DLAs are fundamentally different from those of the millimeter wave accelerators.}
Since each row of dielectric grating \textcolor{black}{in the multi-channel DLA} is identical, this DLA structure is a finite photonic crystal. 
Consistent with Ref. \cite{leedle2018phase}, we focus on silicon pillars and refer this design as a Multi-Input Multi-Output Silicon Accelerator (MIMOSA).

\begin{figure}[!t]
    \centering
    \includegraphics[width=0.45\textwidth]{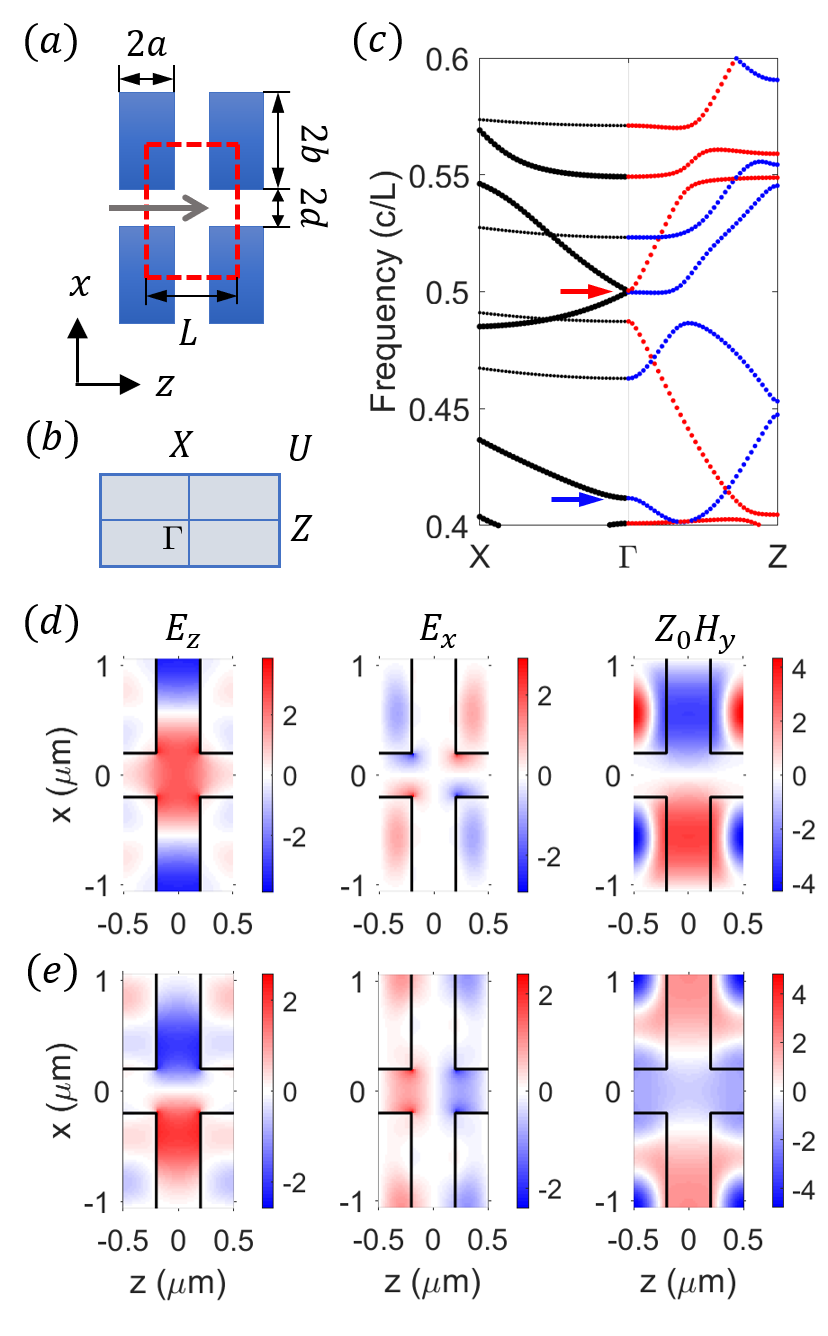}
    \caption{\label{figbandacc} (a) Illustration of the photonic crystal where the unit cell is highlighted in the dashed red box. The electron propagation direction is indicated by the gray arrow. $2a$, $2b$, $2d$ and $L$ respectively represent pillar length, pillar width, gap width, and the periodicity in the electron propagation direction. (b) The reciprocal Brillouin zone. (c) Band diagram of the TM mode of the photonic crystal with $L = 1$ $\mu$m, $a = 0.3$ $\mu$m, $b = 0.86$ $\mu$m and $d = 0.2$ $\mu$m.  Big (small) black dots represent eigenmodes with odd (even) mirror-\textit{z} symmetry, and red (blue) dots represent eigenmodes with even (odd) mirror-\textit{x} symmetry. (d) and (e) are field profiles of the eigen modes at $\Gamma$ point with normalized frequency $0.500$ and $0.411$ respectively. }
\end{figure} 

We find that the essential characteristics of the MIMOSA are captured in the band structure and eigenmode properties of the underlying infinite photonic crystal. For simplicity, we study a two dimensional photonic crystal with a rectangular lattice (Fig. \ref{figbandacc}), where the dielectric pillars extend uniformly in the \textit{y}-direction. Previous studies confirm the validity of using a two-dimensional calculation to describe a photonic crystal with pillar height larger than wavelength \cite{leedle2018phase, black2019laser}. In a two-dimensional photonic crystal, the fields can be decomposed into TE (with $E_y$, $H_x$ and $H_z$ nonzero) and TM (with $E_x$, $E_z$ and $H_y$ nonzero) polarization \cite{joannopoulos2008molding}. Since only the TM polarization provides acceleration, we study only the TM polarization in this work. The unit cell and the band diagram of the photonic crystal underlying the MIMOSA are illustrated in Fig. \ref{figbandacc}(a) and (c) respectively. We assume that the electrons travel along \textit{z}-direction and are centered around $x=0$ and the incident light propagates along the \textit{x}-direction. Therefore, the incident laser can excite eigenmodes lying on $\Gamma X$ direction in the reciprocal space (Fig. \ref{figbandacc}(b)). Furthermore, the mode at $\Gamma$ point has the same phase in each unit cell. Thus,
to ensure that the electron beams in different channels get the same acceleration, the mode at $\Gamma$ point should be excited dominantly.
Therefore, the frequency $\omega$ of the eigenmode at $\Gamma$ point should match the frequency $\omega_0$ of the incident light. In addition, in order to satisfy the phase synchronization condition for DLA \cite{black2019laser, niedermayer2017beam}, we need to have
\begin{equation}
    \label{eqsynchronization}
    \omega(\Gamma) = \omega_0 = \beta \times 2\pi m c /L ,
\end{equation}
where $c$ is the speed of light, $\beta=v/c$ where $v$ is the speed of electron, $L$ is the periodicity along \textit{z}-direction, and $m$ is the diffraction order. In typical DLAs, the amplitude of the first order diffraction is stronger than that of higher order diffraction, so we take $m=1$ in the synchronization condition.

Additionally, the unit cell of typical photonic crystals may have certain symmetries. In the demonstration shown in Fig. \ref{figbandacc}, the mirror-\textit{x} and mirror-\textit{z} symmetries of the unit cell require that the acceleration mode should also have certain symmetries. To accelerate the electron traveling along $x=0$, the mode should be symmetric with respect to \textcolor{black}{the mirror plane} $x=0$ (red dots in Fig. \ref{figbandacc}(c)). And to couple to plane waves propagating in \textit{x}-direction, the mode should have odd mirror-\textit{z} symmetry. 
In other words, the photonic crystal underlying the MIMOSA should support an eigenmode at normalized frequency (frequency normalized by $c/L$) $\beta$ with odd mirror-\textit{z} and even mirror-\textit{x} symmetry, below we refer to such a mode as an acceleration mode. This condition on the symmetry of acceleration mode, together with Eq. (\ref{eqsynchronization}), represents
one of the main contributions of this study and is referred to as the \textit{band structure condition} below. It can be satisfied through tuning the geometrical parameters of the dielectric pillar. 

Moreover, the figure of merit of the MIMOSA can also be derived from eigenmode analysis of the underlying photonic crystal. This ``acceleration factor" $g$ \cite{byer2013development, wei2017dualbragg, Miao2019surface} is the maximal acceleration gradient at the center of the electron channel divided by the maximal electric field inside the dielectric material. 
From the field profile of the acceleration mode (Fig. \ref{figschematic}(d)), which is computed for a structure that is infinitely periodic along the both the \textit{x} and \textit{z}-directions, we can calculate the acceleration factor, which turns out to be close to the acceleration factor of the MIMOSA, where the structure is finite along the \textit{x}-direction. 

Based on the discussions above, the MIMOSA design procedure is summarized as following: (1) Given the electron speed and the frequency of the excitation laser, the periodicity ($L$) of the photonic crystal along the electron propagation direction is determined by Eq. \ref{eqsynchronization}. (2) The periodicity along the transverse direction and the shape of the dielectric pillar in the unit cell can be tuned such that the \textit{band structure condition} is satisfied and the width of the electron channel exceeds a value set by experimental considerations. (3) Using the acceleration factor of the acceleration mode as the objective function, the unit cell of the photonic crystal is optimized with the constraints in (1) and (2). (4) The procedure above involves only the study of an infinitely periodic photonic crystal. After such a photonic crystal is designed,  we can confirm the design by simulating a photonic crystal with a finite number of periods, as we will discuss in Sec. \ref{secdemo}.

\section{Demonstration and analysis}
\label{secdemo}

Using the design procedure as discussed above, we demonstrate the design of a MIMOSA for electron speed 0.5c ($\beta = 0.5$) with rectangular silicon pillars and a central laser wavelength at 2 $\mu$m ($\lambda_0=2$ $\mu$m). The same design principles apply to other electron speeds, pillar shapes, dielectric material systems and laser wavelengths. The periodicity of the photonic crystal is determined by the phase synchronization condition  $L = \beta \lambda_0 = 1$ $\mu$m \cite{black2019laser}. We denote the half length and width of the rectangular pillar as $a$ and $b$ respectively, and half width of the electron channel along \textit{z}-direction as $d$. Larger channel width generally results in higher beam current but lower acceleration gradient due to the exponential decay of the near fields. To compromise between the requirements of high acceleration gradient and wide channel width, we choose $d = 0.2$ $\mu $m, consistent with previous studies \cite{leedle2015dielectric, leedle2018phase}. By tuning $a$, $b$, we find that with parameters $a = 0.3$ $\mu $m and $b = 0.86$ $\mu $m, the photonic crystal supports an acceleration mode at normalized frequency 0.5 (Fig. \ref{figbandacc}(c)) with high acceleration factor.
The mode profile is shown in Fig. \ref{figbandacc}(d), from which the inferred acceleration factor is $g = 0.51$.

\begin{figure}[!htbp]
    \centering
    \includegraphics[width=0.45\textwidth]{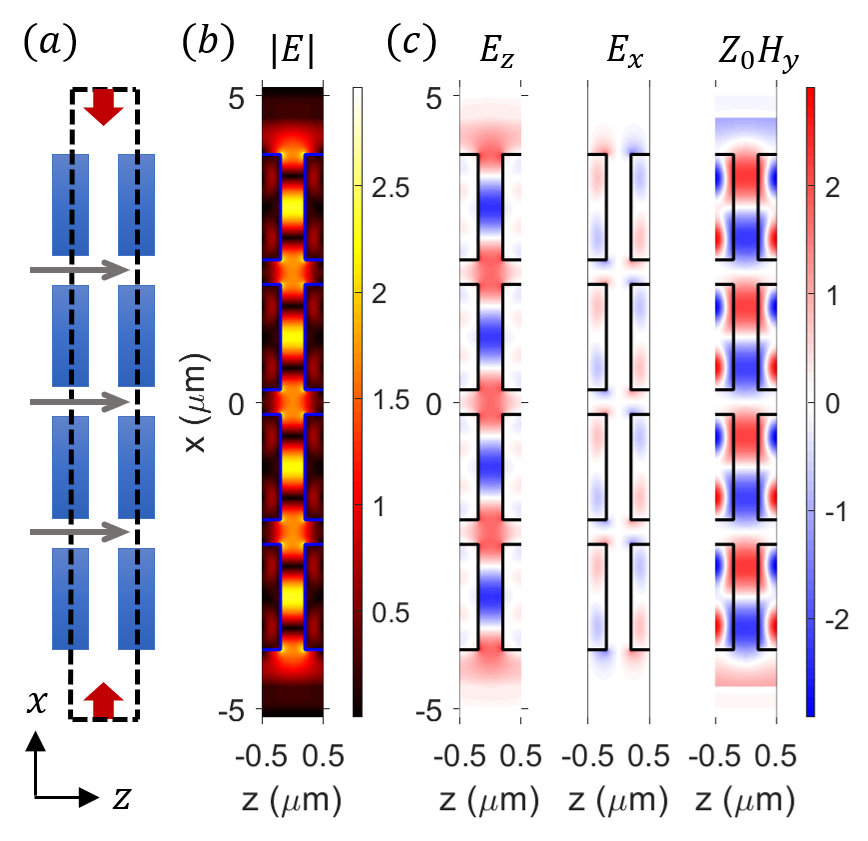}
    \caption{(a) Schematic of the dual drive simulation. The dashed box highlights the unit cell, the red arrows represent the illuminating lasers, and the gray arrows indicate the electron propagation direction. Under in-phase and equal amplitude illumination at wavelength $\lambda_0 = 2$ $\mu$m, the magnitude of E field is shown in (b), while $E_z$, $E_x$, and $Z_0H_y$ are shown in (c), where $Z_0$ is free space impedance.}
    \label{figdualdriveacc}
\end{figure}

To verify our design principles, we truncate the photonic crystal in the \textit{x}-direction to specify a finite number of channels and perform a full wave simulation \cite{shin2012choice}. For demonstration purpose, we limit the number of electron channels to $N=3$. The field distributions are shown in Fig. \ref{figdualdriveacc}, where the two driving plane waves are symmetric with respect to $x=0$. The fields inside different electron channels are almost identical and strongly resemble the eigenmode shown in Fig. \ref{figbandacc}(d), calculated for an infinite photonic crystal. From Fig. \ref{figdualdriveacc}(b), we find that the largest E field is inside the vacuum rather than inside the dielectric, which contributes to the high acceleration factor \cite{Miao2019surface}. This holds when the rectangle pillars have rounded corners with small radius \cite{wei2017dualbragg}.

\begin{figure}[ht]
    \centering
    \includegraphics[width=0.48\textwidth]{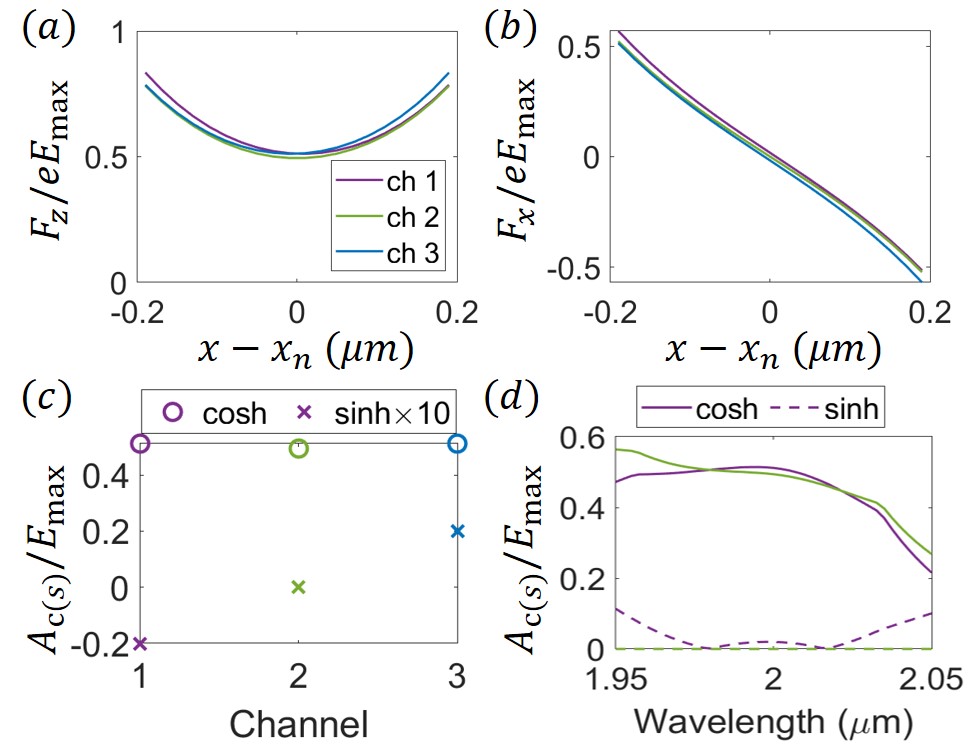}
    \caption{The longitudinal (a) and transverse (b) force distribution inside each electron channel. Amplitudes of the cosh and sinh components in each channel at central frequency (c) and frequency dependence (d). }
    \label{figanalacc}
\end{figure}

When the synchronization condition Eq. \ref{eqsynchronization} is satisfied for $m=1$, and in the limit of a perfectly rigid electron beam, only the first diffraction order interacts with the electron beam \cite{niedermayer2017beam}. The corresponding longitudinal field distribution of the first diffraction order has the following form inside the electron channel \cite{leedle2015dielectric, leedle2018phase}:
\begin{equation}
\begin{split}
    \label{eqEz}
    E_z^n(x,z) =  & \exp(-j2\pi z/L)\{ A_c^n \cosh[\alpha (x-x_n)]\\ & + A_s^n \sinh[\alpha (x-x_n)]\},
\end{split}
\end{equation}
where $A_{c}^n$ and $A_{s}^n$ are the amplitudes of the ``cosh'' and ``sinh'' components, respectively, in the $n^{\textrm{th}}$ electron channel, $\alpha = \sqrt{(2\pi/L)^2 - (\omega/c)^2}$ characterizes the decay of the evanescent wave inside the electron channel, and $x_n$ is the center of the $n^{\textrm{th}}$ channel. $A_c^n$ and $A_s^n$ depend on the incident waves and are discussed in detail in Appendix \ref{app:field}. From Maxwell's equations, we can get $E_x$ and $H_y$. The synchronized Lorentz force $\mathbf{F} = q(\mathbf{E} + \mathbf{v}\times \mathbf{B})$ on the electron is 
\begin{align}
\begin{split}
\label{eqFz}
    F_z(x) = & \, q \textrm{Re}\big\{\exp(j\phi_0) [A_c^n\cosh(\alpha(x-x_n)) \\ & + A_s^n\sinh(\alpha(x-x_n))]\big\}, 
\end{split}\\
\begin{split}
\label{eqFx}
    F_x(x) = & \, q \frac{1}{\gamma} \textrm{Re}\big\{ j\exp(j\phi_0) [A_c^n \sinh(\alpha(x-x_n))\\ & +A_s^n\cosh(\alpha(x-x_n)) ]\big\},
\end{split}
\end{align}
and $F_y(x)=0$, where $q = -e$ is the electron charge, the electron phase $\phi_0 = \omega t_0$ is the phase of the oscillating field experienced by an electron entering the structure at time $t_0$ with respect to the reference particle entering at $t=0$, and $\gamma = 1/\sqrt{1 - \beta^2}$ is the familiar relativistic factor \cite{niedermayer2017beam, black2019laser}. Equations \ref{eqFz} and \ref{eqFx} suggest that the cosh component can provide acceleration (deceleration) or focusing (de-focusing), depending on $\phi_0$, while the sinh component can provide transverse deflection.

Figure \ref{figanalacc}(a) shows the acceleration force $F_z$ where we choose $\phi_0$ such that $F_z$ is maximized at $x=0$. Again choosing $\phi_0$ such that the force is maximized, the transverse force $F_x$ is shown in Fig. \ref{figanalacc}(b), which focuses the electron to the channel center \cite{black2019laser}. The longitudinal forces in different channels differ by less than 7\%, which confirms that the acceleration in different channels are almost identical. 
This holds even better as the number of channels increases, for instance to 10, as shown in Appendix \ref{app:bandwidth}. 

The acceleration factor $g$, derived from the cosh-component of the MIMOSA, is \textcolor{black}{$|A_{c}^n|/E_{\textrm{max}}$} in the $n^{\textrm{th}}$ electron channel, where $E_{\textrm{max}}$ is the maximal electric field inside the dielectric. 
We find the cosh component dominant over the sinh component in each channel. And moreover the cosh components have similar amplitudes in different channels. 
$A_{c}^n/E_{\textrm{max}}=$ 0.512, 0.494 and 0.512 in channels 1, 2, and 3 respectively (Fig. \ref{figanalacc}(c)). These acceleration factors agree with the prediction made by eigenmode analysis, which predicts $g = 0.51$. Due to the mirror-\textit{x} symmetry of the excitation, $A_{c}^n$ ($A_{s}^n$) in channels 1 and 3 are equal in amplitude with same (opposite) sign. Moreover, $A_c^n$ 
are in phase with the symmetric excitation (Appendix \ref{app:field}). Therefore, the electrons with same entering time experience almost identical forces in different channels.

The bandwidth of the MIMOSA is 62 nm, within which the difference between $A_{c}$ in each channel and $A_{c}$ in the central channel at central frequency is less than $10\%$ (Fig. \ref{figanalacc}(d)). Such bandwidth corresponds to 95 fs pulse duration of a transform limited Gaussian pulse at 2 $\mu$m. 
With such a pulse and a fluence half of the damage fluence of silicon (0.17 J/cm$^2$ \cite{lee2017ultrafast}), the predicted acceleration gradient can reach 0.56 GeV/m.
However, as the number of electron channels increases, the bandwidth of the MIMOSA decreases and the corresponding pulse duration increases. We estimate that the corresponding pulse duration scales linearly with the number of channels as $\tau (\textrm{fs}) = 40.8 \times N - 54.4$ for large N (Appendix \ref{app:bandwidth}).

\textcolor{black}{Due to the broadband nature of the MIMOSA, we anticipate the long-range wakefield effects to be insignificant. The short-range wakefield effects and beam loading properties are similar to those in dual grating DLAs \cite{plettner2006proposed, schachter2002optical, siemann2004energy}, and a brief discussion is presented in Appendix \ref{app:beam_loading}. 
The space charge effect in MIMOSA is similar to previous single channel DLAs \cite{england2014dielectric, zhao2018design}, which is briefly discussed in Appendix \ref{app:space_charge}.
Moreover, 
for applications like medical therapy, the required average number of electrons per micro-bunch per channel is only slightly above 1 (Appendix \ref{app:medical}). Therefore, both the space charge and wakefield effects are negligible in the MIMOSA operating for medical applications.}

\textcolor{black}{To match the phase synchronization condition as the electrons get accelerated, we can gradually change the geometric parameters of the MIMOSA unit cell such that the band structure condition (Eq. (\ref{eqsynchronization})) is always satisfied. The tapering of grating periods demonstrated previously \cite{mcneur2018elements, tiberio1998fabrication} can be applied to MIMOSA designs straightforwardly to achieve continuous phase velocity matching. The shape of dielectric pillar can also be tapered such that the frequency of the acceleration mode at $\Gamma$ point continuously matches the central frequency of the driving laser. Furthermore, since the electromagnetic field distributions are almost identical from channel to channel, the alternating phase focusing \cite{niedermayer2018alternating} can also be applied to MIMOSA for long-distance acceleration with stable beam transport.}

\section{Deflector}
\label{sec:deflector}
\begin{figure}
    \centering
    \includegraphics[width=0.45\textwidth]{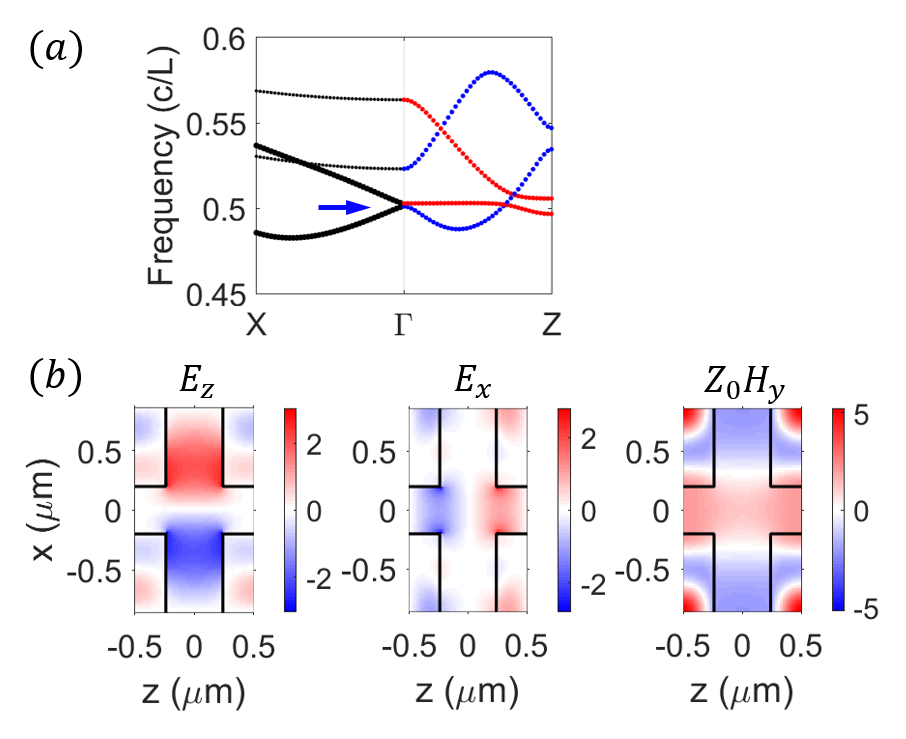}
    \caption{\label{figdualdrivedefl} (a) The band diagram of a photonic crystal deflecting structure with $L = 1$ $\mu$m, $a = 0.26$ $\mu$m, $b = 0.66$ $\mu$m and $d = 0.2$ $\mu$m. The symbols have the same meaning as in Fig. \ref{figbandacc}(c). The field profiles of the deflection mode at $\Gamma$ point with normalized frequency 0.5 (indicated by the blue arrow) are shown in (b). }
\end{figure}

Instead of providing multi-channel acceleration, the MIMOSA can be designed to manipulate electron beams in many other ways. In this section, we consider a MIMOSA designed for simultaneous deflection of N electron beams \cite{leedle2018phase, black2019generation}. The design procedure is almost the same as that of an accelerating-mode MIMOSA. In contrast to the acceleration mode, the deflection mode in the photonic crystal with rectangular pillars has odd mirror-\textit{x} symmetry. The figure of merit of the deflector is the ratio between the deflection gradient at channel center and the maximal electric field inside the dielectric, i.e. $A_s/E_{\textrm{max}}$. 
We start with the photonic crystal shown in Sec. \ref{secdemo} to demonstrate the design procedures (2)-(3).
Although the photonic crystal we show in Sec. \ref{secdemo} supports a deflection eigenmode at frequency around $\omega=0.5 \times 2\pi c/L$, eigenmode analysis shows that $A_s/E_{\textrm{max}}$ is low. Nevertheless, the deflection mode at frequency
$\omega = 0.411\times 2\pi c/L$ has high $A_s/E_{\textrm{max}}$.
To satisfy the \textit{band structure condition}, we tune the geometric parameter of the pillar to $a = 0.26$ $\mu$m and $b = 0.66$ $\mu$m such that the normalized frequency of this deflection mode equals $\beta$ (Fig. \ref{figdualdrivedefl}(a)). We find that the field distributions of the eigenmode remain qualitatively unchanged, as indicated by comparing the deflection mode before (Fig. \ref{figbandacc}(e)) and after (Fig. \ref{figdualdrivedefl}(b)) the parameter tuning.

\begin{figure}
    \centering
    \includegraphics[width=0.9\linewidth]{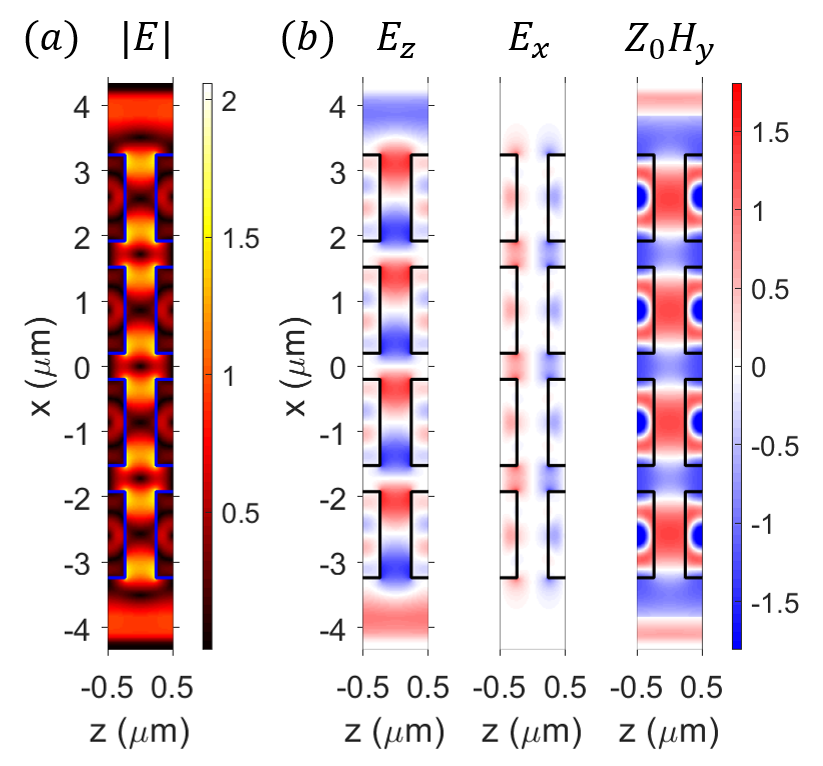}
    \caption{Field distributions in the 3-channel deflecting-mode MIMOSA under anti-symmetric excitation. (a) shows the electric field amplitudes, while (b) shows field components $E_z$, $E_x$ and $Z_0H_y$.}
    \label{fig:defl_field}
\end{figure}

\begin{figure}
    \centering
    \includegraphics[width = 0.96\linewidth]{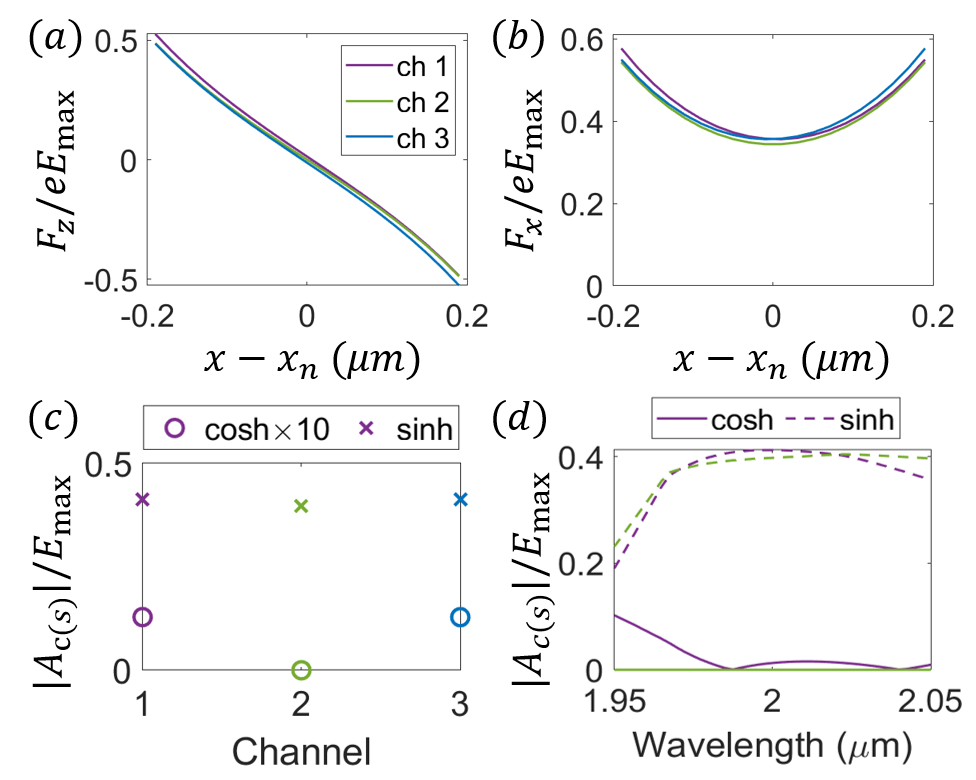}
    \caption{The longitudinal and transverse force distributions in a deflecting-mode MIMOSA are shown in (a) and (b) respectively, with the proper electron phase that maximizes each force. (c) shows the amplitudes of cosh and sinh components in different channels at central frequency, and (d) shows their frequency dependence.}
    \label{figdeflanalysis}
\end{figure}

To validate the design of the photonic crystal electron deflector, we truncate the photonic crystal in \textit{x}-direction to have 3 electron channels and perform a full wave simulation of such finite-width structure. The two driving plane waves propagating in \textit{x}-direction are set to have odd mirror symmetry with respect to $x = 0$. The field distributions are shown in Fig. \ref{fig:defl_field}, which are almost identical from channel to channel. The longitudinal and transverse Lorentz forces calculated from the field distributions are shown in Fig. \ref{figdeflanalysis}(a) and (b) respectively. Since the relative phase between the peaks of longitudinal and transverse forces is $\pi/2$, the longitudinal forces almost vanish when the transverse forces peak.
The deflection force has a cosh-like shape inside the channel, and the channel-to-channel variance is within 7\% (Fig. \ref{figdeflanalysis}(b)).
The amplitudes of cosh and sinh components are shown in Fig. \ref{figdeflanalysis}(c). The fields inside electron channels are dominantly sinh component, where $A_{s}/E_{\textrm{max}} = $ 0.412, 0.398, 0.412 in channel 1, 2, and 3 respectively. Such multi-channel electron deflector also has large bandwidth, as indicated by Fig. \ref{figdeflanalysis}(d). Its bandwidth shares similar scaling rule with the number of electron channels as for accelerating-mode MIMOSAs.
This photonic crystal electron deflector is a natural extension of the laser driven electron deflectors, which are experimentally investigated recently \cite{black2019laser, leedle2018phase}.

\section{Centralizer}
\label{sec:centralizer}

\begin{figure}[!htp]
    \centering
    \includegraphics[width=0.5\linewidth]{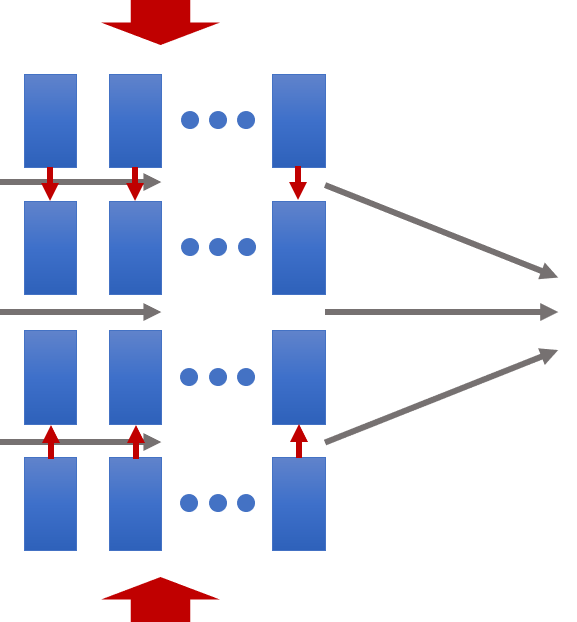}
    \caption{Schematic of a multi-channel centralizer. With symmetric excitation, the transverse forces inside the electrons channels are indicated by the small red arrows. 
    The gray arrows indicate the trajectories of electron beams.}
    \label{fig:centralizer_schematic}
\end{figure}

The MIMOSA can be designed to achieve more complicated functions.
In this section, we further explore its functionalities by studying the modes in MIMOSA with different field distributions from channel to channel. As an example, we demonstrate a 3-channel centralizer which deflects the electron beam on the two outside channels to the central channel (Fig. \ref{fig:centralizer_schematic}). 
This functionality provides the possibility of combining multiple electron beams \textcolor{black}{(super-beam)} to form a high brightness, high current beam \cite{grote2003design}. \textcolor{black}{The centralizer preserves the emittance of the super-beam, while allowing the beamlets travelling in the individual channel to collide at a fixed point. The total emittance is at best a linear addition of the emittance of the individual beamlets.}

\begin{figure}[!htp]
    \centering
    \includegraphics[width = 0.8\linewidth]{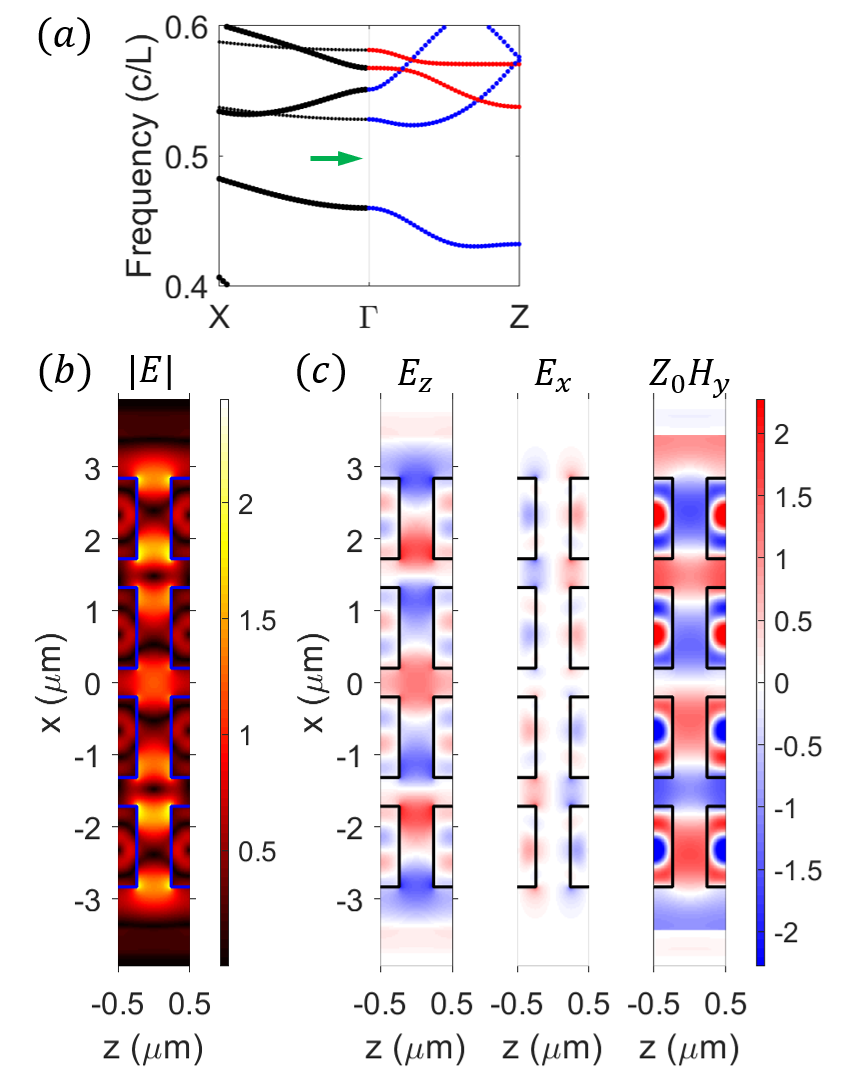}
    \caption{(a) Band structure of the infinitely periodic photonic crystal underlying the electron centralizer. The periodicity in \textit{z} and \textit{x} directions are $L = 1$ $\mu$m and $L_x = 1.52$ $\mu$m, and the dimensions of the rectangular pillar are $a = 0.26$ $\mu$m and $b = 0.56$ $\mu$m. The green arrow indicates that the incident frequency is within the bandgap. With symmetric excitation, the electric field amplitudes are shown in (b), while the non-zero field components $E_z$, $E_x$ and $H_y$ are shown in (c).}
    \label{fighybreigen}
\end{figure}

\begin{figure}[!htp]
    \centering
    \includegraphics[width = 0.96\linewidth]{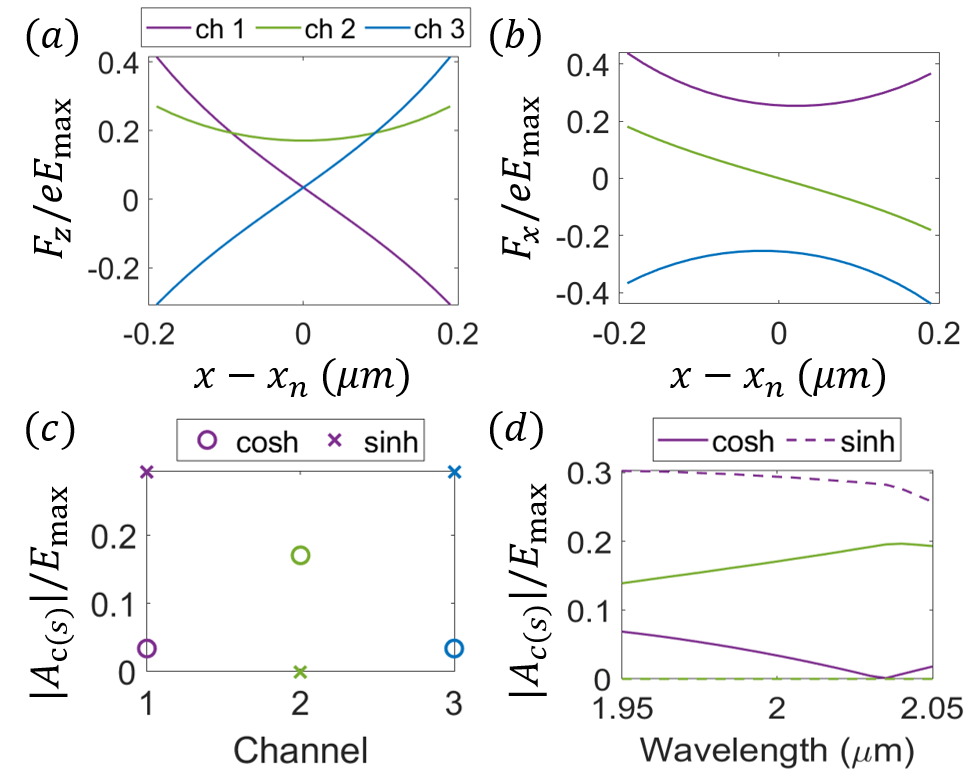}
    \caption{MIMOSA functioning as a centralizer. (a) and (b) show the longitudinal and transverse forces respectively with the electron phases that maximize the longitudinal or transverse forces. The amplitudes of cosh and sinh components are shown in (c), and their frequency dependence is shown in (d).}
    \label{fighybranalysis}
\end{figure}

The underlying photonic crystal is chosen such that the excitation frequency is within the bandgap of the photonic crystal along $\Gamma$X direction. Figure \ref{fighybreigen}(a) shows the band structure of a photonic crystal with $L$ = 1 $\mu$m, $a$ = 0.26 $\mu$m, $b$ = 0.56 $\mu$m, and $d$ = 0.2 $\mu$m. The excitation frequency ($\omega_0 = 0.5\times 2\pi c/L$) is within the bandgap of the photonic crystal.

With symmetric dual drive excitation (Figs. \ref{fighybreigen}(b) and (c)), the field generally have larger magnitudes in the outside channels and smaller magnitudes in the central channel. The force distributions inside different channels are shown in Fig. \ref{fighybranalysis} (a) and (b). The electron phase should be chosen such that the transverse forces are maximized. We find that at this electron phase, the electron beam in the bottom channel is deflected up while the electron beam in the top channel is deflected down. The electron beam travelling along the central channel experiences a small focusing force. Therefore, this device can centralize the electron beams towards the central channel. 
The amplitudes of the cosh and sinh components are shown in Fig. \ref{fighybranalysis}(c). We can find that the fields inside the central channel have only the cosh component while the fields in the two outer channels have dominantly sinh components with equal magnitudes and opposite signs. This centralizer is also broadband, as suggested by the relatively flat frequency dependence shown in Fig. \ref{fighybranalysis}(d).

\section{Discussions and Conclusions}
\label{secconclusion}

The diverse functionalities of MIMOSA, as discussed above, enable sophisticated control of multiple electron beams.
For example, the MIMOSA can provide a platform to study the interference of phase-locked electron beams with the added capability of acceleration, attosecond-scale bunching, and coherent deflection. Furthermore, the interaction between multiple phase-locked electron beams with photonic crystal based radiation generator can potentially boost the radiation generation further \cite{yang2018maximal}.

In conclusion, we here propose a DLA architecture 
based on photonic crystals, that enables simultaneous acceleration of multiple electron beams and has the potential to increase the total beam current by at least one order of magnitude. We find that the characteristics of the MIMOSA can be inferred from the band structure and eigenmodes of the underlying photonic crystal, which provides a simple approach to designing such MIMOSA structures. The underlying photonic crystal should support an eigenmode at $\Gamma$ point with normalized frequency $\beta$. Numerical studies confirm that the field distributions in different channels are indeed almost identical and the acceleration factor is qualitatively consistent with the eigenmode prediction. We further extend the principle to design electron deflectors and other electron manipulation devices based on photonic crystals. Our study opens new opportunities in dielectric laser accelerators and, in general, nanoscale electron manipulation with lasers.

\section*{Acknowledgement}
We wish to acknowledge all contributors in the ACHIP collaboration for their guidance and comments.  This work was supported by the Gordon and Betty Moore Foundation (GBMF4744). Z. Zhao also acknowledges
support from Stanford Graduate Fellowship.

\appendix

\section{Fields in the MIMOSA}
\label{app:field}
In this Appendix, we derive the expression of the electromagnetic fields inside multiple electron channels in the MIMOSA. The derivations partially follow that in \cite{black2019generation}. 

Assuming that the dielectric is nonmagnetic, linear and isotropic, the electric field is a solution to the Maxwell's equations:
\begin{equation}
    \nabla \times \nabla \times \mathbf{E} = \frac{\omega^2}{c^2}\epsilon\mathbf{E} - j\omega \mu_0 \mathbf{J},
\end{equation}
where $\epsilon$ is the relative permittivity, $\mu_0$ is the vacuum permeability, $\mathbf{J}$ is the excitation current, and we consider  the harmonic fields, with $\exp(j\omega t)$ time dependence.
In the quasi-2D approximation, the structure is uniform along \textit{y}-direction and incident plane waves are perpendicular to \textit{y}-direction. Thus, the fields are invariant along \textit{y}-direction and can be decomposed into TM and TE polarization. We further limit our study to the TM polarization (with nonzero $E_x$, $E_z$ and $H_y$ components), which only generates forces in the \textit{xz}-plane. 

The fields inside the photonic crystal satisfy the Bloch theorem when $\mathbf{J}=0$. 
If the 2D photonic crystal is periodic in both \textit{x} and \textit{z} direction, i.e. $\epsilon(\mathbf{r}) = \epsilon(\mathbf{r} + n_x L_x\mathbf{\hat{x}} + n_z L \mathbf{\hat{z}})$ where $n_x$ and $n_z$ are arbitrary integers, $L_x$ and $L$ are periodicity in \textit{x} and \textit{z} directions respectively, the fields satisfy
\begin{equation}
    \label{eqE2D}
    \mathbf{E}(\mathbf{r})=\mathbf{U}(\mathbf{r})\exp(-jk_x x -jk_z z),
\end{equation}
where the periodic part $U(\mathbf{r}) = U(\mathbf{r} + n_x L_x\mathbf{\hat{x}} + n_z L \mathbf{\hat{z}})$, and $k_x$, $k_z$ are Bloch wave vectors in \textit{x} and \textit{z} directions respectively. 

The MIMOSA generally has finite number of channels and can be regarded as a 2D photonic crystal truncated in the \textit{x}-direction. Thus, in the MIMOSA, which is finite in \textit{x}-direction and periodic along \textit{z}-direction, i.e. $\epsilon(\mathbf{r}) = \epsilon(\mathbf{r} + L\hat{z})$, the fields have the following form:
\begin{equation}
    \mathbf{E}(\mathbf{r})=\mathbf{E}_p(\mathbf{r})\exp(-jk_z z),
\end{equation}
where the periodic field $\mathbf{E}_p(\mathbf{r}) = \mathbf{E}_p(\mathbf{r}+L\hat{z})$. The periodic $\mathbf{E}_p$ can be further expanded into a Fourier series,
\begin{equation}
    \mathbf{E}_p(\mathbf{r})=\sum_{m=-\infty}^{m=\infty}\mathbf{e}_m(x)\exp(-jk_m z),
\end{equation}
where $k_m = 2\pi m/L$. Suppose only $m^{\textrm{th}}$ order diffraction is phase synchronized with the electron beam, i.e.
\begin{equation}
    \frac{\omega}{\beta c} = k_m + k_z = \frac{2\pi m}{L} + k_z.
\end{equation}
The forces generated by all other diffraction orders on the electron are averaged over one period to be zero. The synchronized electric fields have the form:
\begin{equation}
    \mathbf{E}_m(\mathbf{r}) = \begin{pmatrix} e_{m,x}(x) \\0 \\ e_{m,z}(x) \end{pmatrix}\exp(-jk_m z - jk_z z),
\end{equation}
where $e_{m,x}$ and $e_{m,z}$ are the phasor notations.
In the following derivation, we consider only this diffraction order and drop the diffraction order index $m$. 

The synchronized field (phasor) inside the $n^{\textrm{th}}$ electron channel has the following form generally,
\begin{align}
\begin{split}
    e_z^n(x) & = d_ne^{-\alpha (x-x_n)} + c_n e^{\alpha (x-x_n)} , \\
    e_x^n(x)&  = \frac{k_m+k_z}{j\alpha}[  d_n e^{-\alpha (x-x_n)} - c_n e^{\alpha (x-x_n)}]
    \end{split}
\end{align}
where $\alpha = \sqrt{(k_m+k_z)^2 - (\omega/c)^2}$,  $d_n$ and $c_n$ are amplitudes of the evanescent waves decaying away from the dielectric on two sides of the electron channel, and $x_n$ is the center of the $n^{\textrm{th}}$ channel. Suppose the periodicity along \textit{x}-direction is $L_x$ and the center of MIMOSA is at $x=0$, we find $x_n = [n - (N+1)/2]L_x$. Due to the velocity of electron being lower than speed of light ($\beta < 1$), the wave vector $k_m + k_z = \omega/\beta c > \omega/c$, and $\alpha$ is always real. This indicates that the synchronized field inside the electron channel is always evanescent and $\alpha$ characterizes the decay of the evanescent wave inside the electron channel. We can reformulate the phasor with hyperbolic cosine and sine functions:
\begin{align}
\label{eqezchannel}
\begin{split}
    e_z^n(x) = \, & A_c^n\cosh[\alpha (x-x_n)] + A_s^n \sinh[\alpha (x-x_n)],\\
    e_x^n(x) = \, &  \frac{k_m+k_z}{j\alpha}\{ A_s^n \cosh[\alpha(x-x_n)] \\ & + A_c^n \sinh[\alpha(x-x_n)]\}
    \end{split}
\end{align}
where $A_c^n = d_n + c_n$ and $A_s^n = d_n - c_n$ are the amplitudes of the cosh and sinh components.

\textit{Case 1}: If the MIMOSA is excited by a plane wave polarized in \textit{z}-direction and propagating in \textit{x}-direction:
\begin{equation}
    \mathbf{E}_{inc}(\mathbf{r}, t) = E_0 \exp(j\omega t-jk_0x)\mathbf{\hat{z}},
\end{equation}
where $k_0 = \omega/c$, the eigenmodes of the photonic crystal with $k_z = 0$ and frequency $\omega$ can be excited. Suppose that only two eigenmodes with frequency $\omega$ and Bloch wave vector ($\pm k_x$, $k_z=0$) are excited. The field inside the truncated photonic crystal can be approximated by a sum of these two counter-propagating Bloch waves, which has the following form:
\begin{align}
\label{eqEBloch}
\begin{split}
    \mathbf{E}(x,z) =  & a_+ \mathbf{U}(x,z; k_x, k_z=0) \exp(-j k_x x) \\ & + a_- \mathbf{U}(x, z; -k_x, k_z=0) \exp(jk_x x),
    \end{split}
\end{align}
where $a_+$ and $a_-$ are amplitudes of the two counter-propagating Bloch waves (Eq. \ref{eqE2D}). 
Thus, the synchronized field (phasor) should have the form:
\begin{align}
    \label{eqeBloch}
    \begin{split}
    \mathbf{e}_1(x) = \, & a_+\mathbf{e}_p(x;k_x) \exp(-j k_x x)\\ & + a_- \mathbf{e}_p(x; -k_x)\exp(jk_x x),
    \end{split}
\end{align}
where the subscript 1 highlights that this is the phasor for \textit{Case 1}, and the periodic part $\mathbf{e}_p(x; \pm k_x) = \mathbf{e}_p(x + L_x; \pm k_x)$. 

Comparing Eq. (\ref{eqezchannel}) and (\ref{eqeBloch}), we find that $A_c^n$ and $A_s^n$ in \textit{Case 1} should have the form:
\begin{align}
\label{eqAc1}
    A_{c,1}^n & = a_+^c\exp(-j k_x x_n) + a_-^c\exp(jk_x x_n), \\
    \label{eqAs1}
    A_{s,1}^n & = a_+^s\exp(-j k_x x_n) + a_-^s\exp(jk_x x_n),
\end{align}
where the coefficients $a_+^c$, $a_-^c$, $a_+^s$, and $a_-^s$ are independent of $n$ and have the following form:
\begin{align}
\label{eqacoeff}
    \begin{split}
    a_+^c & = a_+ e_{p,z}(-\frac{mod(N+1,2)}{2}L_x; k_x), \\
    a_-^c & = a_- e_{p.z}(-\frac{mod(N+1,2)}{2}L_x; -k_x),\\
    a_+^s & = \frac{j\alpha}{k_m}a_+ e_{p,x}(-\frac{mod(N+1,2)}{2}L_x; k_x),\\
    a_-^s & = \frac{j\alpha}{k_m}a_- e_{p,x}(-\frac{mod(N+1,2)}{2}L_x; -k_x).
    \end{split}
\end{align}

From Eqs. \ref{eqAc1} and \ref{eqAs1}, we find that generally the amplitudes of cosh and sinh components ($A_{c,1}^n$ and $A_{s,1}^n$) are different from channel to channel in terms of both magnitudes and phases. However, when $k_x = 0$, the amplitudes of cosh and sinh components in different electron channels become identical. This implies that even with single side drive, it is possible to achieve identical acceleration in multiple channels of the MIMOSA. Nevertheless, the condition $k_x = 0$ is usually satisfied only at a single frequency. For broadband excitation ($\sim$ 300 fs laser pulse \cite{leedle2018phase}), the dual drive can selectively excite the desired mode and in general have larger bandwidth.

\textit{Case 2}: If the MIMOSA is excited by a $z$-polarized plane wave propagating in $-x$-direction, i.e. $\mathbf{E}_{inc}(\mathbf{r},t) = E_0 \exp(j\omega t + jk_0 x)\mathbf{\hat{z}}$, the field is a mirror image of the field in \textit{Case 1} due to the mirror-$x$ symmetry of MIMOSA. 
\begin{align}
    \begin{split}
    \mathbf{e}_2(x) = \, & M_x[\mathbf{e}_1(-x)]\\  = \, & a_- M_x[\mathbf{e}_p(-x; -k_x)] \exp(-jk_x x)\\
    & + a_+ M_x[\mathbf{e}_p(-x; k_x)] \exp(jk_x x),
    \end{split}
\end{align}
where $M_x$ represents a mirror-$x$ operation on the vector such that 
\begin{equation}
    M_x\begin{bmatrix} e_x \\ e_y \\ e_z \end{bmatrix} = \begin{bmatrix} -e_x \\ e_y \\ e_z \end{bmatrix}.
\end{equation}
Similar to \textit{Case 1}, $A_c^n$ and $A_s^n$ in \textit{Case 2} are
\begin{align}
\label{eqAc2}
    A_{c,2}^n & = a_-^c \exp(-jk_x x_n) + a_+^c \exp(jk_x x_n), \\
    \label{eqAs2}
    A_{s,2}^n & = -a_-^s \exp(-jk_x x_n) - a_+^s \exp(jk_x x_n).
\end{align}
Comparing Eqs. \ref{eqAc1}, \ref{eqAs1}, \ref{eqAc2}, and \ref{eqAs2}, and recall that $x_n = [n - (N+1)/2]L_x = - x_{N+1-n}$, we find that 
\begin{align}
\label{eqA1A2}
    \begin{split}
        A_{c,2}^n & = A_{c,1}^{N+1-n}, \\
        A_{s,2}^n & = - A_{s,1}^{N+1-n}.
    \end{split}
\end{align}

\textit{Case 3}: If the MIMOSA is excited by two counter propagating plane waves in \textit{x}-direction with equal amplitudes and relative phase $\theta$, i.e. $ \mathbf{E}_{inc}(\mathbf{r},t) = E_0 [\exp(-jk_0 x) + \exp(j\theta) \exp(jk_0 x)] \exp(j\omega t)\mathbf{\hat{z}}$, the field in this case is a sum of the field in \textit{Case 1} and \textit{Case 2} with relative phase $\theta$.
\begin{align}
    \mathbf{e}_3(x) = \mathbf{e}_1(x) + \exp(j\theta) \mathbf{e}_2(x)
\end{align}
Therefore, the amplitudes of cosh and sinh components in the $n^{\textrm{th}}$ electron channel are:
\begin{align}
\begin{split}
    A_{c,3}^n = \, & A_{c,1}^n + \exp(j\theta) A_{c,2}^n \\
    = \, & a_+^c [\exp(-j k_x x_n) + \exp(j\theta) \exp(jk_x x_n)] \\
    & + a_-^c[ \exp(jk_x x_n) + \exp(j\theta)\exp(-j k_x x_n)],
\end{split}\\
\begin{split}
    A_{s,3}^n = \, & A_{s,1}^n + \exp(j\theta) A_{s,2}^n \\ 
    = \, & a_+^s[\exp(-jk_x x_n) - \exp(j\theta) \exp(j k_x x_n)] \\
    & + a_-^s[\exp(jk_x x_n) - \exp(j\theta) \exp(-jk_x x_n)].
\end{split}
\end{align}

With symmetric excitation, i.e. $\theta = 0$, the amplitudes of cosh and sinh components are
\begin{align}
\label{eqA3sym}
    \begin{split}
    A_{c,3}^n & = 2(a_+^c + a_-^c) \cos(k_x x_n), \\
    A_{s,3}^n & = -2j (a_+^s - a_-^s)\sin(k_x x_n).
    \end{split}
\end{align}
From Eq. \ref{eqA3sym}, we observe that the channel-to-channel variance in the amplitudes of cosh and sinh components is minimized when $k_x$ is near zero. Therefore, to provide near identical force for electrons traveling in different channels, the MIMOSA should operate near the $\Gamma$ point ($k_x=0$). When $k_x$ is near zero, either real if the excitation frequency is within the band or purely imaginary if the excitation frequency is within the band gap, the phases of $A_{c,3}^n$ are identical for different channels. 
These are observed in the numerical study. We emphasize that these properties are crucial to provide near identical acceleration to micro-bunched electrons with the same entering time.

Similarly, with anti-symmetric excitation, i.e. $\theta = \pi$, the amplitudes of cosh and sinh components in different electron channels are
\begin{align}
\label{eqA3antisym}
    \begin{split}
    A_{c,3}^n & = -2j(a_+^c - a_-^c) \sin(k_x x_n), \\
    A_{s,3}^n & = 2 (a_+^s + a_-^s)\cos(k_x x_n).
    \end{split}
\end{align}
The amplitudes of sinh components also have the same phase and small variation in magnitude for $k_x$ near zero.

Since the studied MIMOSA have mirror-$x$ symmetry, Eq. \ref{eqA1A2} implies that $A_{c,3}^n$ ($A_{s,3}^n$) are symmetric (anti-symmetric) with respect to $x=0$ under symmetric excitation, while $A_{c,3}^n$ ($A_{s,3}^n$) are anti-symmetric (symmetric) with respect to $x=0$ under anti-symmetric excitation. These are consistent with Eqs. \ref{eqA3sym} and \ref{eqA3antisym}. 

We also find that a MIMOSA with high acceleration factor may not have high deflection factor when the dual drive excitation changes from symmetric to anti-symmetric. From Eqs. \ref{eqA3sym} and \ref{eqA3antisym}, it is obvious that the cosh component amplitudes under symmetric excitation are not necessarily the same as the sinh component amplitudes under anti-symmetric excitation. This observation is aligned with the analysis in \cite{black2019generation}.

\section{The MIMOSA with large number of electron channels}
\label{app:bandwidth}

\begin{figure}[!htp]
    \centering
    \includegraphics[width = 0.98\linewidth]{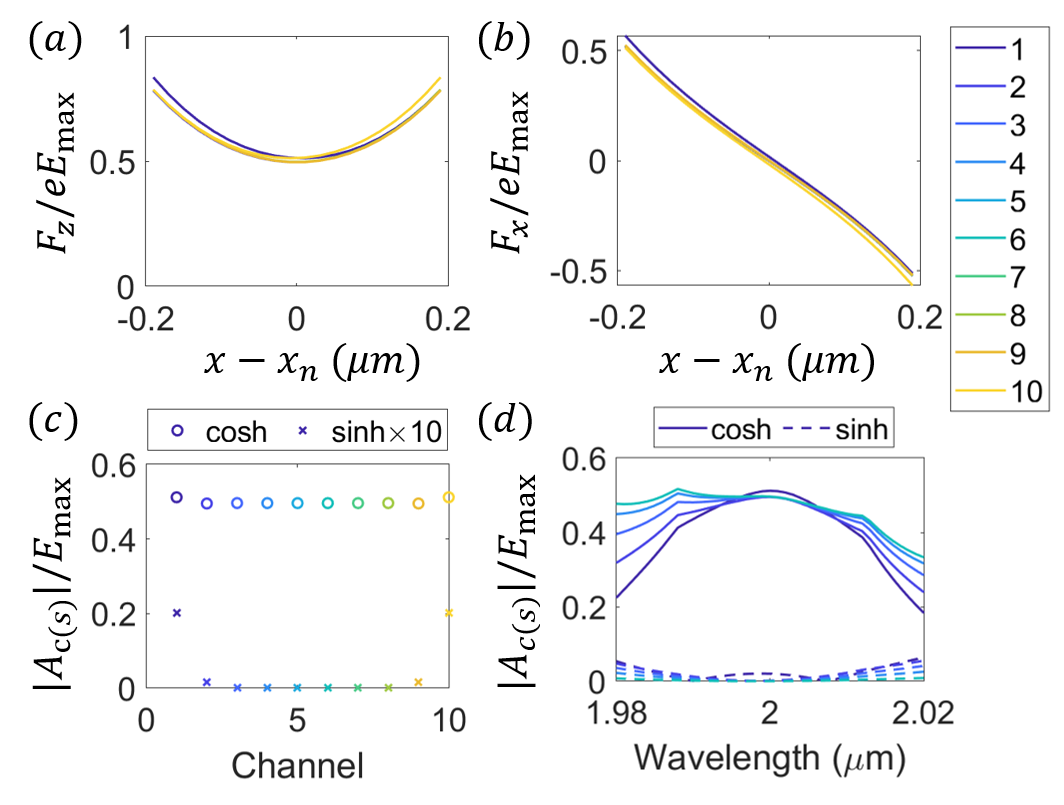}
    \caption{MIMOSA with 10 electron channels. (a) and (b) show the longitudinal and transverse forces in different channels. The amplitudes of cosh and sinh components in different channels are shown in (c), and their wavelength dependence is shown in (d). Different colors indicate different channels. Due to the mirror-$x$ symmetry, channels above $x=0$ are omitted in (d).}
    \label{figchannel10}
\end{figure}

In this Appendix, we discuss the figure of merits of MIMOSA with large number of electron channels and the scaling rule of its bandwidth.
In the main text, we demonstrate a MIMOSA with 3 electron channels. Here, we extend the number of electron channels to 10 with all other parameters fixed. With symmetric excitation and the proper electron phase to maximize the acceleration force, the longitudinal force distributions inside different channels are shown in Fig. \ref{figchannel10}(a). With another electron phase that maximizes the focusing force, the transverse force distributions inside different channels are shown in Fig. \ref{figchannel10}(b). The almost overlapping curves imply that the force distributions are almost identical from channel to channel. 
Indeed, the amplitudes of cosh components in different channels differ by less than 3\%, and the amplitude of cosh components dominate over that of sinh components (Fig. \ref{figchannel10}(c)).

\begin{figure}[!htp]
    \centering
    \includegraphics[width = 0.6\linewidth]{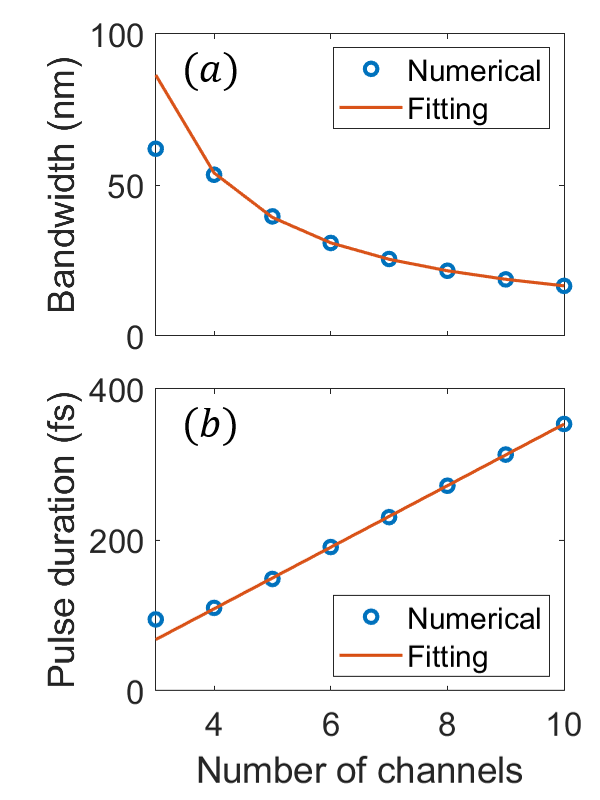}
    \caption{(a) Bandwidth of the MIMOSA versus number of electron channels. The geometric parameters are the same as those studied in Sec. III of the main text. (b) The corresponding pulse duration of a transform limited Gaussian pulse with central wavelength 2 $\mu$m and a bandwidth matching the bandwidth of the MIMOSA.}
    \label{figbandwidth}
\end{figure}

Nevertheless, with increasing number of channels, the bandwidth of the MIMOSA decreases. From Fig. \ref{figchannel10}(d), we find that the bandwidth decreases to 16.6 nm in the 10-channel MIMOSA. The bandwidth approximately scales with the inverse of the number of pillars, since the MIMOSA can be regarded as an optical resonator where the stored energy scales linearly with the number of pillars while the energy leakage rate remains roughly constant. 
For the studied MIMOSA, the bandwidth ($\Delta \lambda$), in nm,  as a function of number of channels ($N$) is estimated as $\Delta \lambda (\text{nm}) =  144/(N - 1.33)$ when $N \geq 4$, as shown in Fig. \ref{figbandwidth}(a). 
If we choose a transform limited Gaussian pulse with central wavelength at 2 $\mu$m and bandwidth matching the bandwidth of the MIMOSA, the relation between the pulse duration ($\tau$) and the number of channels is linear (Fig. \ref{figbandwidth}(b)). The estimated scaling rule is $\tau(
\textrm{fs}) = 40.8 \times N - 54.4$ for $N \geq 4$, where $\tau$ is the full width at half maximum pulse duration in femto-second.
Therefore, due to the bandwidth scaling rule, the number of electron channels cannot be arbitrarily large and should be chosen to match the bandwidth of the excitation pulse.

We study further the scaling coefficient in the relation between the matched pulse duration ($\tau$) and the number of channels ($N$), which is generally linear for large $N$,
\begin{equation}
    \label{eqbandwidthscale}
    \tau = \alpha_1 N + \alpha_2,
\end{equation}
where the scaling coefficient $\alpha_1$ determines how fast the bandwidth narrows as the number of channels increases. 
To achieve broadband operation and large number of channels simultaneously, a small scaling coefficient is favorable.
We find that this scaling coefficient depends on the photonic crystal and is related to the band structure of the underlying photonic crystal along $\Gamma X$ direction, specifically, related to the band containing the acceleration mode and the band gap adjacent to the acceleration mode.
When the band is close to a flat band, i.e. the band has small curvature near $\Gamma$ point, a small change in the excitation frequency leads to large change in $k_x$ of the mode, which results in narrow bandwidth. On the other hand, when the excitation frequency is within the band gap near the acceleration mode, field decays exponentially inside the MIMOSA from the out-most pillars, which increases the field variation from channel to channel. Since the penetration depth is positively correlated to the band curvature near $\Gamma$ point \cite{zhao2019penetration}, a large band curvature, or even a linear dispersion relation, near the $\Gamma$ point is favorable. Therefore, to achieve small scaling coefficient $\alpha_1$, the underlying photonic crystal should have large band curvature, ideally a linear dispersion relation, near the $\Gamma$ point. We take this design rule into account when we optimize the geometric parameters of the MIMOSA. As shown in Fig. \ref{figbandacc}(c) and \ref{figdualdrivedefl}(a) of the main text, the bands near the acceleration mode (red arrow in Fig. \ref{figbandacc}(c)) and the deflection mode (blue arrow in Fig. \ref{figdualdrivedefl}(a)) have nearly linear dispersion relation.

\begin{figure}
    \centering
    \includegraphics[width = 0.85\linewidth]{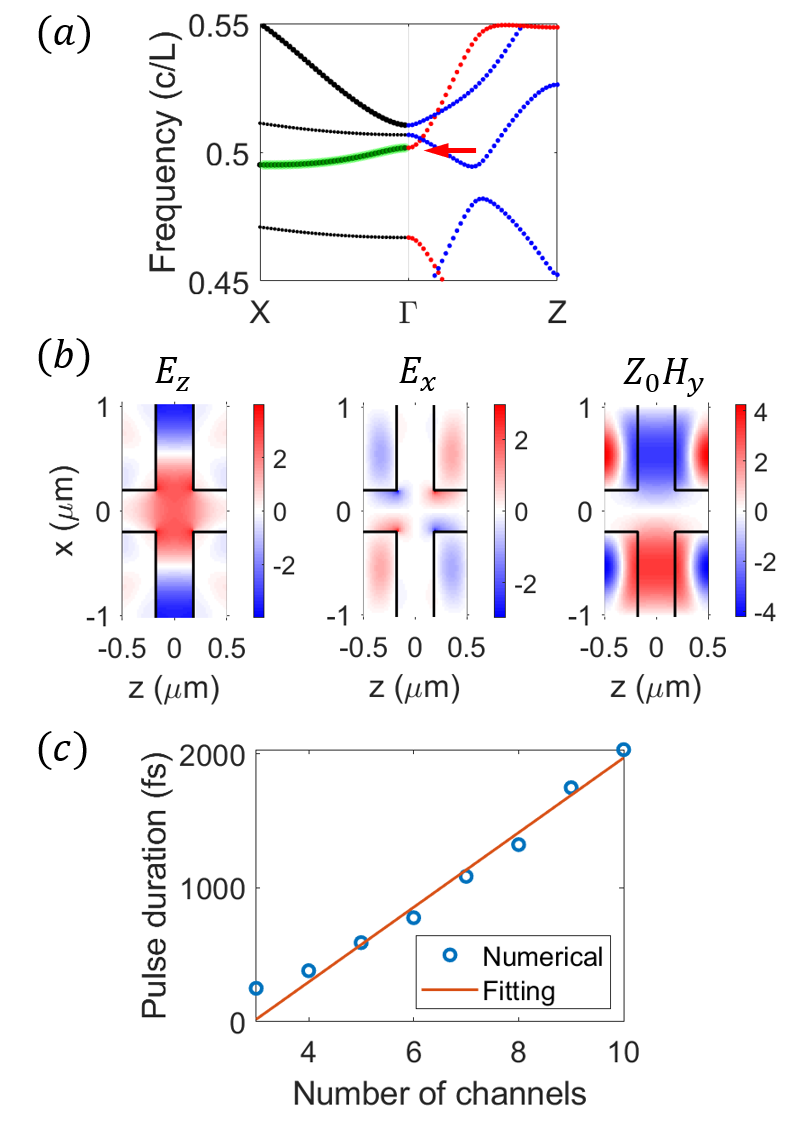} 
    \caption{MIMOSA with high acceleration factor but small bandwidth. (a) The band structure of the photonic crystal ($L=1$ $\mu$m, $a=0.32$ $\mu$m, $b=0.82$ $\mu$m, $d = 0.2$ $\mu$m). The red arrow points to the acceleration mode. The band containing the acceleration mode is highlighted in green. (b) The field distribution of the acceleration mode. (c) The pulse duration of the transform limited Gaussian pulse, whose bandwidth matches the bandwidth of the MIMOSA, as a function of the number of channels.}
    \label{fignarrowbandwidth}
\end{figure}

To emphasize the importance of small bandwidth scaling factor in a MIMOSA, We show a counter-example of a MIMOSA with high acceleration factor but narrow bandwidth due to the small band curvature. As shown in Fig. \ref{fignarrowbandwidth}(a), the photonic crystal with rectangle pillars ($L=1$ $\mu$m, $a=0.32$ $\mu$m, $b=0.82$ $\mu$m, $d = 0.2$ $\mu$m) support an acceleration mode with normalized frequency around 0.5. From the field distribution of the acceleration mode (Fig. \ref{fignarrowbandwidth}(b)), the acceleration factor is as high as 0.53. However, the band containing the acceleration mode has small curvature, as highlighted in green in Fig. \ref{fignarrowbandwidth}(a). As a result, the bandwidth decreases fast as the number of channels increases. Figure \ref{fignarrowbandwidth}(c) shows the pulse duration of the transform limited Gaussian pulse with the same bandwidth as the MIMOSA. We estimate the scaling is $\tau(\textrm{fs}) = 279 \times N - 819$, where the scaling coefficient $\alpha_1=279$ is much larger than that of the MIMOSA shown in Fig. \ref{figbandacc} of the main text ($\alpha_1 = 40.8$).

\section{Requirements on electron bunches for medical applications}
\label{app:medical}

\textcolor{black}{The linear accelerators used for medical therapy usually operate on a dose rate around 1 $\textrm{Gy}\cdot\textrm{min}^{-1}$ \cite{sousa2009dose}. If we assume a typical phantom mass of 1 kilogram and a typical electron beam final energy of 6 MeV, the required average beam current is 2.8 nA. Supposing the femto-second laser operates at 50 MHz repetition rate with 250 fs pulse duration \cite{black2019laser, leedle2018phase}, the corresponding number of electrons per laser pulse is $3.5\times10^2$. Recent progress in laser-driven nanotip electron sources has demonstrated this required number of electrons per laser pulse, with below $\textrm{nm}\cdot\textrm{rad}$ emittance \cite{tafel2019femtosecond}. From the bandwidth analysis in Appendix \ref{app:bandwidth} (Fig. \ref{figbandwidth}), we find that a 7-channel MIMOSA matches this pulse duration. If these electrons are distributed over the 7 channels and 37 micro-bunches, which falls within the pulse duration of 250 fs, the average number of electrons per micro-bunch per channel is only 1.3.}

\textcolor{black}{To summarize, for medical applications, the required number of electrons per laser pulse is low due to the high laser repetition rate. With the multi-channel DLA architecture and the micro-bunching technology, the required number of electrons per micro-bunch per channel is around 1.}

\section{Beam loading properties of MIMOSA}
\label{app:beam_loading}

\textcolor{black}{In this Appendix, we provide a brief discussion of the short-range wakefield effects and beam loading properties of MIMOSA, although a comprehensive study of the wakefields in such multi-channel DLA is beyond the scope of this study due to the intrinsic complexity of Cheronkov radiation in photonic crystals \cite{luo2003cerenkov}.}

\textcolor{black}{To study the beam loading properties of the MIMOSA, we follow the discussion presented in \cite{plettner2006proposed}. Suppose the unloaded gradient is $G_0$, the loaded gradient is $G_L=G_0-Nqk_W-qk_H$, where $N$ is the number of electron channels, $q$ is the charge per channel per micro-bunch, $k_W$ represents the loss factor due to the particle’s wakefield that overlaps with the excitation laser field, and $k_H$ represents the nonoverlapping component of the wakefields such as the broadband Cherenkov radiation. We assume that the micro-bunches in different channels are in phase. Thus, their wakefield components that overlap with the laser field will add up coherently, which gives rise to the multiple of $N$ in the expression of the loaded gradient. Although the Cherenkov radiation from the in-phase micro-bunches is also coherent at a particular frequency, the average effect over the broadband resembles an incoherent sum, due to their constructive interference over some frequencies and destructive interference over some other frequencies.  }

\textcolor{black}{From \cite{plettner2006proposed, schachter2002optical}, we find that $k_W=\frac{cZ_C}{4\lambda^2\beta}$, where $Z_C$ is the characteristic impedance of the MIMOSA, and $k_H=\frac{c}{\lambda^2}Z_H$, where $Z_H$ is the broadband wakefield impedance. The characteristic impedance is $Z_C=\xi^2 \frac{\beta \lambda}{D} Z_0$, where $Z_0=\sqrt{\mu_0/\epsilon_0 }$ is the impedance of vacuum, $\xi$ is the ratio of the unloaded gradient and the incident electric field, i.e. $\xi=G_0/E_0 \sim 0.5$, and D is the extension in the out-of-plane dimension. The broadband wakefield impedance depends on the Cherenkov radiation in the MIMOSA structure, which is intrinsically complicated and not discussed in detail in this study. The wakefield impedance can be approximated by a wide bunch, with extension $D$ in the out-of-plane direction, traveling between two gratings with separation $2d$. The impedance is approximately $Z_H=\frac{\lambda^2}{2π R_\textrm{eff}^2} Z_0$ and $R_\textrm{eff}= \sqrt{dD}$. Using the properties of the MIMOSA design presented in Section \ref{secdemo} and assuming $D=\lambda=2$ $μ$m, the characteristic impedance is $Z_C=47$ $\Omega$ and the broadband wakefield impedance is $Z_H=600$ $\Omega$. }

\textcolor{black}{For the 7-channel MIMOSA operating for medical applications discussed in Appendix \ref{app:medical}, the reduction in acceleration gradient due to wakefields is 12 kV/m, much smaller than the unloaded gradient ($\sim 0.25 \textrm{GV/m}$). This observation confirms our anticipation that the wakefield effects are small in MIMOSAs for medical applications.}

\textcolor{black}{We also provide an estimation of the energy efficiency of MIMOSA and the optimal charge per bunch that achieve the maximal energy efficiency. The energy efficiency is $\eta=\frac{NqG_L L}{P_L \tau_{\textrm{pulse}}}$, where $L$ is the periodicity along the propagation direction, $P_L=LDE_0^2/Z_0$ is the laser power incident on one period from two sides, and $\tau_\textrm{pulse}$ is the pulse duration. }

\textcolor{black}{Consistent with discussions in Appendix \ref{app:medical}, we assume the laser pulse duration is $\tau_\textrm{pulse}=250$ fs, and take $N=7$ to match this pulse duration. For the MIMOSA operating for medical applications, $q = 1.3e$ and the energy efficiency is $1.1\times 10^{-6}$. Even a train of 37 bunches are fed in the laser pulse, the energy efficiency is only $4.1\times10^{-5}$. Such low energy efficiency suggest the demand of recycling the laser pulses. }

\textcolor{black}{Nevertheless, the energy efficiency of MIMOSA can be improved significantly with a proper choice of charge per micro-bunch. To maximize the efficiency, the optimal charge is $q_\textrm{opt}=\frac{G_0}{2(Nk_W+k_H)}$ \cite{plettner2006proposed, siemann2004energy} and the optimal efficiency is $\eta_\textrm{opt}=\frac{1}{1 + 4\beta Z_H/(NZ_C)} \frac{\tau_c}{\tau_\textrm{pulse}}$, where $\tau_c$ is the duration of an optical cycle. With the aforementioned impedance of the MIMOSA design, we find that $q_\textrm{opt}=2.18$ fC, which corresponds to $1.36\times10^4$ electrons per micro-bunch, and $\eta_\textrm{opt}=0.58\%$. Furthermore, if a train of micro-bunches is fed into the MIMOSA structure, the efficiency can be increased further. For a train of 37 bunches, which falls within the pulse duration of 250 fs, the efficiency becomes 21\%, under the approximation that the long-range interaction of the wake fields is negligible due to the low resonance nature of MIMOSA.}

\section{Space charge effects in MIMOSA}
\label{app:space_charge}

\begin{figure}[t]
    \centering
    \includegraphics[width=0.96\linewidth]{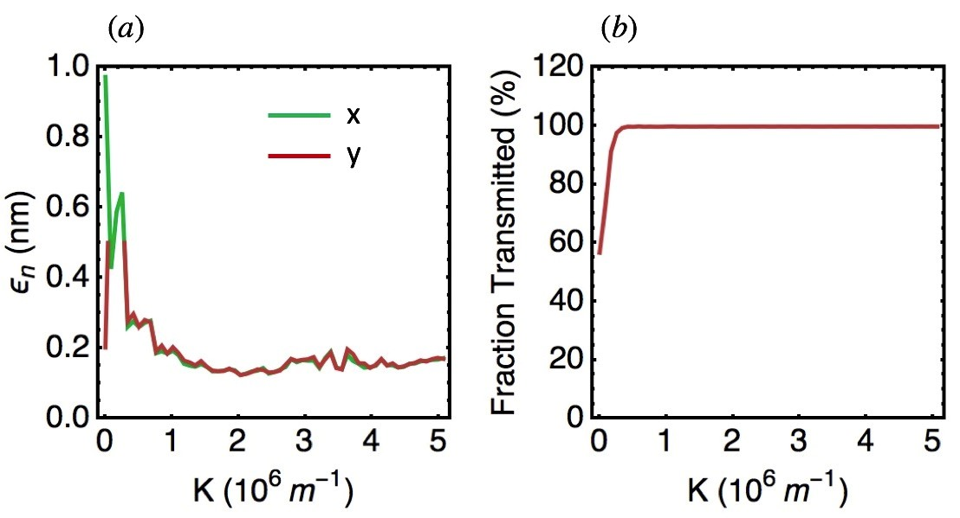}
    \caption{\textcolor{black}{Particle tracking simulations of 500 attosecond FWHM bunch with initial emittance of 0.1 nm and charge of 2 fC showing (a) final emittance after propagating through the MIMOSA structure and (b) corresponding fraction of transmitted particles as function of externally applied focusing field K. Only the center channel of the MIMOSA is shown since the results for the three channels are nearly identical. }}
    \label{fig:emittance_focus}
\end{figure}

\textcolor{black}{In this Appendix, we discuss the space charge effects of an attosecond electron bunch injected into the MIMOSA and the required external focusing field to avoid emittance growing. }

\textcolor{black}{An ideal scenario for beam coupling to the MIMOSA architecture is a line of point emitters spaced at the channel separation triggered in unison by a single laser pulse. Such a linear array of electron sources can be produced using nanotip field emitters etched in silicon, as discussed in \cite{ceballos2014silicon}.  Consistent with recent DLA experiments \cite{black2019generation, Miao2019surface}, the 3-channel MIMOSA has 15 periods in the electron propagation direction, whose geometric parameters are given in Sec. \ref{secdemo}.  We use the particle tracking code General Particle Tracer (GPT) to simulate the normalized transverse emittance of a short electron bunch under the influence of space-charge effects in this scenario. For the tracking simulation, we assume that each channel has incident on it a microbunch of duration 500 attoseconds full-width at half maximum (FWHM), consistent with recently demonstrated optically bunched beams \cite{sears2008production, black2019laser, schonenberger2019generation}. A transverse normalized emittance of 0.1 nm is assumed with a bunch charge of 2 fC, consistent with the estimated optimal bunch charge for efficient beam loading calculated in Appendix \ref{app:beam_loading}.}

\textcolor{black}{Due to a combination of space charge repulsion, emittance pressure, and transverse defocusing in the narrow accelerating channel, strong transverse focusing forces are required for beam confinement. The normalized focusing force K may be interpreted as the linear focusing term appearing in a z-dependent paraxial ray equation of the form $x^{\prime \prime}(z) = - K^2 x(z)$. We note that the minimal focusing strength for optimal emittance preservation and near 100\% particle transport in Fig. \ref{fig:emittance_focus} ($K^2$ = 4 $\times 10^{12}$ m$^{-2}$) is of the order of magnitude estimated in Ref. \cite{england2014dielectric} as being required for focusing of sub-relativistic electron beams in DLA structures. For the present tracking simulation we use a solenoidal magnetic field $B$ oriented in the z-direction, which provides a transverse focusing force $K = e B / 2 m c \beta \gamma$. The minimum emittance point in Fig. \ref{fig:emittance_focus}(a) corresponds to a field of $B$ = 3.8 kT, which is impractically large. However, compatible laser driven focusing techniques capable of forces of this magnitude, which have been proposed in \cite{naranjo2012stable, niedermayer2018alternating} and recently demonstrated in \cite{black2019laser}, can be readily adapted to the MIMOSA architecture. However, a full implementation of such a ponderomotive focusing scheme lies beyond the scope of the present paper. Although these results were calculated for all three channels, in Fig. \ref{fig:emittance_focus} we show only the result for the center channel of the MIMOSA device. Since the electromagnetic design successfully equalizes the accelerating fields in the channels, the results for the other 2 channels are nearly identical and displaying all three on one plot would provide no additional information. Also of note is that the uncompensated emittance growth (at $K$ = 0) in Fig. \ref{fig:emittance_focus} (a) is significantly larger in the $x$ coordinate due to the fact that the dominant transverse defocusing is in this dimension, as predicted by Eq. (\ref{eqFx}) and Fig. \ref{figanalacc}. }

\bibliography{mimosa.bib}

\begin{thebibliography}{45}%
\makeatletter
\providecommand \@ifxundefined [1]{%
 \@ifx{#1\undefined}
}%
\providecommand \@ifnum [1]{%
 \ifnum #1\expandafter \@firstoftwo
 \else \expandafter \@secondoftwo
 \fi
}%
\providecommand \@ifx [1]{%
 \ifx #1\expandafter \@firstoftwo
 \else \expandafter \@secondoftwo
 \fi
}%
\providecommand \natexlab [1]{#1}%
\providecommand \enquote  [1]{``#1''}%
\providecommand \bibnamefont  [1]{#1}%
\providecommand \bibfnamefont [1]{#1}%
\providecommand \citenamefont [1]{#1}%
\providecommand \href@noop [0]{\@secondoftwo}%
\providecommand \href [0]{\begingroup \@sanitize@url \@href}%
\providecommand \@href[1]{\@@startlink{#1}\@@href}%
\providecommand \@@href[1]{\endgroup#1\@@endlink}%
\providecommand \@sanitize@url [0]{\catcode `\\12\catcode `\$12\catcode
  `\&12\catcode `\#12\catcode `\^12\catcode `\_12\catcode `\%12\relax}%
\providecommand \@@startlink[1]{}%
\providecommand \@@endlink[0]{}%
\providecommand \url  [0]{\begingroup\@sanitize@url \@url }%
\providecommand \@url [1]{\endgroup\@href {#1}{\urlprefix }}%
\providecommand \urlprefix  [0]{URL }%
\providecommand \Eprint [0]{\href }%
\providecommand \doibase [0]{https://doi.org/}%
\providecommand \selectlanguage [0]{\@gobble}%
\providecommand \bibinfo  [0]{\@secondoftwo}%
\providecommand \bibfield  [0]{\@secondoftwo}%
\providecommand \translation [1]{[#1]}%
\providecommand \BibitemOpen [0]{}%
\providecommand \bibitemStop [0]{}%
\providecommand \bibitemNoStop [0]{.\EOS\space}%
\providecommand \EOS [0]{\spacefactor3000\relax}%
\providecommand \BibitemShut  [1]{\csname bibitem#1\endcsname}%
\let\auto@bib@innerbib\@empty
\bibitem [{\citenamefont {Plettner}\ \emph {et~al.}(2006)\citenamefont
  {Plettner}, \citenamefont {Lu},\ and\ \citenamefont
  {Byer}}]{plettner2006proposed}%
  \BibitemOpen
  \bibfield  {author} {\bibinfo {author} {\bibfnamefont {T.}~\bibnamefont
  {Plettner}}, \bibinfo {author} {\bibfnamefont {P.}~\bibnamefont {Lu}}, and\
  \bibinfo {author} {\bibfnamefont {R.}~\bibnamefont {Byer}},\ }\bibfield
  {title} {\bibinfo {title} {Proposed few-optical cycle laser-driven particle
  accelerator structure},\ }\href@noop {} {\bibfield  {journal} {\bibinfo
  {journal} {Physical Review Special Topics-Accelerators and Beams}\ }\textbf
  {\bibinfo {volume} {9}},\ \bibinfo {pages} {111301} (\bibinfo {year}
  {2006})}\BibitemShut {NoStop}%
\bibitem [{\citenamefont {England}\ \emph {et~al.}(2014)\citenamefont
  {England}, \citenamefont {Noble}, \citenamefont {Bane}, \citenamefont
  {Dowell}, \citenamefont {Ng}, \citenamefont {Spencer}, \citenamefont
  {Tantawi}, \citenamefont {Wu}, \citenamefont {Byer}, \citenamefont {Peralta},
  \citenamefont {Soong}, \citenamefont {Chang}, \citenamefont {Montazeri},
  \citenamefont {Wolf}, \citenamefont {Cowan}, \citenamefont {Dawson},
  \citenamefont {Gai}, \citenamefont {Hommelhoff}, \citenamefont {Huang},
  \citenamefont {Jing}, \citenamefont {McGuinness}, \citenamefont {Palmer},
  \citenamefont {Naranjo}, \citenamefont {Rosenzweig}, \citenamefont {Travish},
  \citenamefont {Mizrahi}, \citenamefont {Schachter}, \citenamefont {Sears},
  \citenamefont {Werner},\ and\ \citenamefont {Yoder}}]{england2014dielectric}%
  \BibitemOpen
  \bibfield  {author} {\bibinfo {author} {\bibfnamefont {R.~J.}\ \bibnamefont
  {England}}, \bibinfo {author} {\bibfnamefont {R.~J.}\ \bibnamefont {Noble}},
  \bibinfo {author} {\bibfnamefont {K.}~\bibnamefont {Bane}}, \bibinfo {author}
  {\bibfnamefont {D.~H.}\ \bibnamefont {Dowell}}, \bibinfo {author}
  {\bibfnamefont {C.-K.}\ \bibnamefont {Ng}}, \bibinfo {author} {\bibfnamefont
  {J.~E.}\ \bibnamefont {Spencer}}, \bibinfo {author} {\bibfnamefont
  {S.}~\bibnamefont {Tantawi}}, \bibinfo {author} {\bibfnamefont
  {Z.}~\bibnamefont {Wu}}, \bibinfo {author} {\bibfnamefont {R.~L.}\
  \bibnamefont {Byer}}, \bibinfo {author} {\bibfnamefont {E.}~\bibnamefont
  {Peralta}}, \bibinfo {author} {\bibfnamefont {K.}~\bibnamefont {Soong}},
  \bibinfo {author} {\bibfnamefont {C.-M.}\ \bibnamefont {Chang}}, \bibinfo
  {author} {\bibfnamefont {B.}~\bibnamefont {Montazeri}}, \bibinfo {author}
  {\bibfnamefont {S.~J.}\ \bibnamefont {Wolf}}, \bibinfo {author}
  {\bibfnamefont {B.}~\bibnamefont {Cowan}}, \bibinfo {author} {\bibfnamefont
  {J.}~\bibnamefont {Dawson}}, \bibinfo {author} {\bibfnamefont
  {W.}~\bibnamefont {Gai}}, \bibinfo {author} {\bibfnamefont {P.}~\bibnamefont
  {Hommelhoff}}, \bibinfo {author} {\bibfnamefont {Y.-C.}\ \bibnamefont
  {Huang}}, \bibinfo {author} {\bibfnamefont {C.}~\bibnamefont {Jing}},
  \bibinfo {author} {\bibfnamefont {C.}~\bibnamefont {McGuinness}}, \bibinfo
  {author} {\bibfnamefont {R.~B.}\ \bibnamefont {Palmer}}, \bibinfo {author}
  {\bibfnamefont {B.}~\bibnamefont {Naranjo}}, \bibinfo {author} {\bibfnamefont
  {J.}~\bibnamefont {Rosenzweig}}, \bibinfo {author} {\bibfnamefont
  {G.}~\bibnamefont {Travish}}, \bibinfo {author} {\bibfnamefont
  {A.}~\bibnamefont {Mizrahi}}, \bibinfo {author} {\bibfnamefont
  {L.}~\bibnamefont {Schachter}}, \bibinfo {author} {\bibfnamefont
  {C.}~\bibnamefont {Sears}}, \bibinfo {author} {\bibfnamefont {G.~R.}\
  \bibnamefont {Werner}}, and\ \bibinfo {author} {\bibfnamefont {R.~B.}\
  \bibnamefont {Yoder}},\ }\bibfield  {title} {\bibinfo {title} {Dielectric
  laser accelerators},\ }\href@noop {} {\bibfield  {journal} {\bibinfo
  {journal} {Reviews of Modern Physics}\ }\textbf {\bibinfo {volume} {86}},\
  \bibinfo {pages} {1337} (\bibinfo {year} {2014})}\BibitemShut {NoStop}%
\bibitem [{\citenamefont {Peralta}\ \emph {et~al.}(2013)\citenamefont
  {Peralta}, \citenamefont {Soong}, \citenamefont {England}, \citenamefont
  {Colby}, \citenamefont {Wu}, \citenamefont {Montazeri}, \citenamefont
  {McGuinness}, \citenamefont {McNeur}, \citenamefont {Leedle}, \citenamefont
  {Walz} \emph {et~al.}}]{peralta2013demonstration}%
  \BibitemOpen
  \bibfield  {author} {\bibinfo {author} {\bibfnamefont {E.}~\bibnamefont
  {Peralta}}, \bibinfo {author} {\bibfnamefont {K.}~\bibnamefont {Soong}},
  \bibinfo {author} {\bibfnamefont {R.}~\bibnamefont {England}}, \bibinfo
  {author} {\bibfnamefont {E.}~\bibnamefont {Colby}}, \bibinfo {author}
  {\bibfnamefont {Z.}~\bibnamefont {Wu}}, \bibinfo {author} {\bibfnamefont
  {B.}~\bibnamefont {Montazeri}}, \bibinfo {author} {\bibfnamefont
  {C.}~\bibnamefont {McGuinness}}, \bibinfo {author} {\bibfnamefont
  {J.}~\bibnamefont {McNeur}}, \bibinfo {author} {\bibfnamefont
  {K.}~\bibnamefont {Leedle}}, \bibinfo {author} {\bibfnamefont
  {D.}~\bibnamefont {Walz}},  \emph {et~al.},\ }\bibfield  {title} {\bibinfo
  {title} {Demonstration of electron acceleration in a laser-driven dielectric
  microstructure},\ }\href@noop {} {\bibfield  {journal} {\bibinfo  {journal}
  {Nature}\ }\textbf {\bibinfo {volume} {503}},\ \bibinfo {pages} {91}
  (\bibinfo {year} {2013})}\BibitemShut {NoStop}%
\bibitem [{\citenamefont {Breuer}\ and\ \citenamefont
  {Hommelhoff}(2013)}]{breuer2013laser}%
  \BibitemOpen
  \bibfield  {author} {\bibinfo {author} {\bibfnamefont {J.}~\bibnamefont
  {Breuer}}and\ \bibinfo {author} {\bibfnamefont {P.}~\bibnamefont
  {Hommelhoff}},\ }\bibfield  {title} {\bibinfo {title} {Laser-based
  acceleration of nonrelativistic electrons at a dielectric structure},\
  }\href@noop {} {\bibfield  {journal} {\bibinfo  {journal} {Physical Review
  Letters}\ }\textbf {\bibinfo {volume} {111}},\ \bibinfo {pages} {134803}
  (\bibinfo {year} {2013})}\BibitemShut {NoStop}%
\bibitem [{\citenamefont {Breuer}\ \emph {et~al.}(2014)\citenamefont {Breuer},
  \citenamefont {McNeur},\ and\ \citenamefont
  {Hommelhoff}}]{breuer2014dielectric}%
  \BibitemOpen
  \bibfield  {author} {\bibinfo {author} {\bibfnamefont {J.}~\bibnamefont
  {Breuer}}, \bibinfo {author} {\bibfnamefont {J.}~\bibnamefont {McNeur}}, and\
  \bibinfo {author} {\bibfnamefont {P.}~\bibnamefont {Hommelhoff}},\ }\bibfield
   {title} {\bibinfo {title} {Dielectric laser acceleration of electrons in the
  vicinity of single and double grating structures--theory and simulations},\
  }\href@noop {} {\bibfield  {journal} {\bibinfo  {journal} {Journal of Physics
  B: Atomic, Molecular and Optical Physics}\ }\textbf {\bibinfo {volume}
  {47}},\ \bibinfo {pages} {234004} (\bibinfo {year} {2014})}\BibitemShut
  {NoStop}%
\bibitem [{\citenamefont {Cesar}\ \emph
  {et~al.}(2018{\natexlab{a}})\citenamefont {Cesar}, \citenamefont {Maxson},
  \citenamefont {Shen}, \citenamefont {Wootton}, \citenamefont {Tan},
  \citenamefont {England},\ and\ \citenamefont {Musumeci}}]{cesar2018enhanced}%
  \BibitemOpen
  \bibfield  {author} {\bibinfo {author} {\bibfnamefont {D.}~\bibnamefont
  {Cesar}}, \bibinfo {author} {\bibfnamefont {J.}~\bibnamefont {Maxson}},
  \bibinfo {author} {\bibfnamefont {X.}~\bibnamefont {Shen}}, \bibinfo {author}
  {\bibfnamefont {K.}~\bibnamefont {Wootton}}, \bibinfo {author} {\bibfnamefont
  {S.}~\bibnamefont {Tan}}, \bibinfo {author} {\bibfnamefont {R.}~\bibnamefont
  {England}}, and\ \bibinfo {author} {\bibfnamefont {P.}~\bibnamefont
  {Musumeci}},\ }\bibfield  {title} {\bibinfo {title} {Enhanced energy gain in
  a dielectric laser accelerator using a tilted pulse front laser},\
  }\href@noop {} {\bibfield  {journal} {\bibinfo  {journal} {Optics Express}\
  }\textbf {\bibinfo {volume} {26}},\ \bibinfo {pages} {29216} (\bibinfo {year}
  {2018}{\natexlab{a}})}\BibitemShut {NoStop}%
\bibitem [{\citenamefont {Cesar}\ \emph
  {et~al.}(2018{\natexlab{b}})\citenamefont {Cesar}, \citenamefont {Custodio},
  \citenamefont {Maxson}, \citenamefont {Musumeci}, \citenamefont {Shen},
  \citenamefont {Threlkeld}, \citenamefont {England}, \citenamefont {Hanuka},
  \citenamefont {Makasyuk}, \citenamefont {Peralta} \emph
  {et~al.}}]{cesar2018high}%
  \BibitemOpen
  \bibfield  {author} {\bibinfo {author} {\bibfnamefont {D.}~\bibnamefont
  {Cesar}}, \bibinfo {author} {\bibfnamefont {S.}~\bibnamefont {Custodio}},
  \bibinfo {author} {\bibfnamefont {J.}~\bibnamefont {Maxson}}, \bibinfo
  {author} {\bibfnamefont {P.}~\bibnamefont {Musumeci}}, \bibinfo {author}
  {\bibfnamefont {X.}~\bibnamefont {Shen}}, \bibinfo {author} {\bibfnamefont
  {E.}~\bibnamefont {Threlkeld}}, \bibinfo {author} {\bibfnamefont
  {R.}~\bibnamefont {England}}, \bibinfo {author} {\bibfnamefont
  {A.}~\bibnamefont {Hanuka}}, \bibinfo {author} {\bibfnamefont
  {I.}~\bibnamefont {Makasyuk}}, \bibinfo {author} {\bibfnamefont
  {E.}~\bibnamefont {Peralta}},  \emph {et~al.},\ }\bibfield  {title} {\bibinfo
  {title} {High-field nonlinear optical response and phase control in a
  dielectric laser accelerator},\ }\href@noop {} {\bibfield  {journal}
  {\bibinfo  {journal} {Communications Physics}\ }\textbf {\bibinfo {volume}
  {1}},\ \bibinfo {pages} {46} (\bibinfo {year}
  {2018}{\natexlab{b}})}\BibitemShut {NoStop}%
\bibitem [{\citenamefont {Wootton}\ \emph
  {et~al.}(2016{\natexlab{a}})\citenamefont {Wootton}, \citenamefont {Wu},
  \citenamefont {Cowan}, \citenamefont {Hanuka}, \citenamefont {Makasyuk},
  \citenamefont {Peralta}, \citenamefont {Soong}, \citenamefont {Byer},\ and\
  \citenamefont {England}}]{wootton2016demonstration}%
  \BibitemOpen
  \bibfield  {author} {\bibinfo {author} {\bibfnamefont {K.~P.}\ \bibnamefont
  {Wootton}}, \bibinfo {author} {\bibfnamefont {Z.}~\bibnamefont {Wu}},
  \bibinfo {author} {\bibfnamefont {B.~M.}\ \bibnamefont {Cowan}}, \bibinfo
  {author} {\bibfnamefont {A.}~\bibnamefont {Hanuka}}, \bibinfo {author}
  {\bibfnamefont {I.~V.}\ \bibnamefont {Makasyuk}}, \bibinfo {author}
  {\bibfnamefont {E.~A.}\ \bibnamefont {Peralta}}, \bibinfo {author}
  {\bibfnamefont {K.}~\bibnamefont {Soong}}, \bibinfo {author} {\bibfnamefont
  {R.~L.}\ \bibnamefont {Byer}}, and\ \bibinfo {author} {\bibfnamefont {R.~J.}\
  \bibnamefont {England}},\ }\bibfield  {title} {\bibinfo {title}
  {Demonstration of acceleration of relativistic electrons at a dielectric
  microstructure using femtosecond laser pulses},\ }\href@noop {} {\bibfield
  {journal} {\bibinfo  {journal} {Optics Letters}\ }\textbf {\bibinfo {volume}
  {41}},\ \bibinfo {pages} {2696} (\bibinfo {year}
  {2016}{\natexlab{a}})}\BibitemShut {NoStop}%
\bibitem [{\citenamefont {Leedle}\ \emph {et~al.}(2015)\citenamefont {Leedle},
  \citenamefont {Ceballos}, \citenamefont {Deng}, \citenamefont {Solgaard},
  \citenamefont {Pease}, \citenamefont {Byer},\ and\ \citenamefont
  {Harris}}]{leedle2015dielectric}%
  \BibitemOpen
  \bibfield  {author} {\bibinfo {author} {\bibfnamefont {K.~J.}\ \bibnamefont
  {Leedle}}, \bibinfo {author} {\bibfnamefont {A.}~\bibnamefont {Ceballos}},
  \bibinfo {author} {\bibfnamefont {H.}~\bibnamefont {Deng}}, \bibinfo {author}
  {\bibfnamefont {O.}~\bibnamefont {Solgaard}}, \bibinfo {author}
  {\bibfnamefont {R.~F.}\ \bibnamefont {Pease}}, \bibinfo {author}
  {\bibfnamefont {R.~L.}\ \bibnamefont {Byer}}, and\ \bibinfo {author}
  {\bibfnamefont {J.~S.}\ \bibnamefont {Harris}},\ }\bibfield  {title}
  {\bibinfo {title} {Dielectric laser acceleration of sub-100 kev electrons
  with silicon dual-pillar grating structures},\ }\href@noop {} {\bibfield
  {journal} {\bibinfo  {journal} {Optics Letters}\ }\textbf {\bibinfo {volume}
  {40}},\ \bibinfo {pages} {4344} (\bibinfo {year} {2015})}\BibitemShut
  {NoStop}%
\bibitem [{\citenamefont {Hughes}\ \emph {et~al.}(2017)\citenamefont {Hughes},
  \citenamefont {Veronis}, \citenamefont {Wootton}, \citenamefont {England},\
  and\ \citenamefont {Fan}}]{hughes2017method}%
  \BibitemOpen
  \bibfield  {author} {\bibinfo {author} {\bibfnamefont {T.}~\bibnamefont
  {Hughes}}, \bibinfo {author} {\bibfnamefont {G.}~\bibnamefont {Veronis}},
  \bibinfo {author} {\bibfnamefont {K.~P.}\ \bibnamefont {Wootton}}, \bibinfo
  {author} {\bibfnamefont {R.~J.}\ \bibnamefont {England}}, and\ \bibinfo
  {author} {\bibfnamefont {S.}~\bibnamefont {Fan}},\ }\bibfield  {title}
  {\bibinfo {title} {Method for computationally efficient design of dielectric
  laser accelerator structures},\ }\href@noop {} {\bibfield  {journal}
  {\bibinfo  {journal} {Optics Express}\ }\textbf {\bibinfo {volume} {25}},\
  \bibinfo {pages} {15414} (\bibinfo {year} {2017})}\BibitemShut {NoStop}%
\bibitem [{\citenamefont {Hughes}\ \emph {et~al.}(2018)\citenamefont {Hughes},
  \citenamefont {Tan}, \citenamefont {Zhao}, \citenamefont {Sapra},
  \citenamefont {Leedle}, \citenamefont {Deng}, \citenamefont {Miao},
  \citenamefont {Black}, \citenamefont {Solgaard}, \citenamefont {Harris},
  \citenamefont {Vuckovic}, \citenamefont {Byer}, \citenamefont {Fan},
  \citenamefont {England}, \citenamefont {Lee},\ and\ \citenamefont
  {Qi}}]{hughes2017chip}%
  \BibitemOpen
  \bibfield  {author} {\bibinfo {author} {\bibfnamefont {T.~W.}\ \bibnamefont
  {Hughes}}, \bibinfo {author} {\bibfnamefont {S.}~\bibnamefont {Tan}},
  \bibinfo {author} {\bibfnamefont {Z.}~\bibnamefont {Zhao}}, \bibinfo {author}
  {\bibfnamefont {N.~V.}\ \bibnamefont {Sapra}}, \bibinfo {author}
  {\bibfnamefont {K.~J.}\ \bibnamefont {Leedle}}, \bibinfo {author}
  {\bibfnamefont {H.}~\bibnamefont {Deng}}, \bibinfo {author} {\bibfnamefont
  {Y.}~\bibnamefont {Miao}}, \bibinfo {author} {\bibfnamefont {D.~S.}\
  \bibnamefont {Black}}, \bibinfo {author} {\bibfnamefont {O.}~\bibnamefont
  {Solgaard}}, \bibinfo {author} {\bibfnamefont {J.~S.}\ \bibnamefont
  {Harris}}, \bibinfo {author} {\bibfnamefont {J.}~\bibnamefont {Vuckovic}},
  \bibinfo {author} {\bibfnamefont {R.~L.}\ \bibnamefont {Byer}}, \bibinfo
  {author} {\bibfnamefont {S.}~\bibnamefont {Fan}}, \bibinfo {author}
  {\bibfnamefont {R.~J.}\ \bibnamefont {England}}, \bibinfo {author}
  {\bibfnamefont {Y.~J.}\ \bibnamefont {Lee}}, and\ \bibinfo {author}
  {\bibfnamefont {M.}~\bibnamefont {Qi}},\ }\bibfield  {title} {\bibinfo
  {title} {On-chip laser-power delivery system for dielectric laser
  accelerators},\ }\href@noop {} {\bibfield  {journal} {\bibinfo  {journal}
  {Physical Review Applied}\ }\textbf {\bibinfo {volume} {9}},\ \bibinfo
  {pages} {054017} (\bibinfo {year} {2018})}\BibitemShut {NoStop}%
\bibitem [{\citenamefont {Hughes}\ \emph {et~al.}(2019)\citenamefont {Hughes},
  \citenamefont {England},\ and\ \citenamefont
  {Fan}}]{hughes2019reconfigurable}%
  \BibitemOpen
  \bibfield  {author} {\bibinfo {author} {\bibfnamefont {T.~W.}\ \bibnamefont
  {Hughes}}, \bibinfo {author} {\bibfnamefont {R.~J.}\ \bibnamefont {England}},
  and\ \bibinfo {author} {\bibfnamefont {S.}~\bibnamefont {Fan}},\ }\bibfield
  {title} {\bibinfo {title} {Reconfigurable photonic circuit for controlled
  power delivery to laser-driven accelerators on a chip},\ }\href
  {https://doi.org/10.1103/PhysRevApplied.11.064014} {\bibfield  {journal}
  {\bibinfo  {journal} {Physical Review Applied}\ }\textbf {\bibinfo {volume}
  {11}},\ \bibinfo {pages} {064014} (\bibinfo {year} {2019})}\BibitemShut
  {NoStop}%
\bibitem [{\citenamefont {Tan}\ \emph {et~al.}(2019)\citenamefont {Tan},
  \citenamefont {Zhao}, \citenamefont {Urbanek}, \citenamefont {Hughes},
  \citenamefont {Lee}, \citenamefont {Fan}, \citenamefont {Harris},\ and\
  \citenamefont {Byer}}]{tan2019silicon}%
  \BibitemOpen
  \bibfield  {author} {\bibinfo {author} {\bibfnamefont {S.}~\bibnamefont
  {Tan}}, \bibinfo {author} {\bibfnamefont {Z.}~\bibnamefont {Zhao}}, \bibinfo
  {author} {\bibfnamefont {K.}~\bibnamefont {Urbanek}}, \bibinfo {author}
  {\bibfnamefont {T.}~\bibnamefont {Hughes}}, \bibinfo {author} {\bibfnamefont
  {Y.~J.}\ \bibnamefont {Lee}}, \bibinfo {author} {\bibfnamefont
  {S.}~\bibnamefont {Fan}}, \bibinfo {author} {\bibfnamefont {J.~S.}\
  \bibnamefont {Harris}}, and\ \bibinfo {author} {\bibfnamefont {R.~L.}\
  \bibnamefont {Byer}},\ }\bibfield  {title} {\bibinfo {title} {Silicon nitride
  waveguide as a power delivery component for on-chip dielectric laser
  accelerators},\ }\href@noop {} {\bibfield  {journal} {\bibinfo  {journal}
  {Optics Letters}\ }\textbf {\bibinfo {volume} {44}},\ \bibinfo {pages} {335}
  (\bibinfo {year} {2019})}\BibitemShut {NoStop}%
\bibitem [{\citenamefont {Zhao}\ \emph {et~al.}(2018)\citenamefont {Zhao},
  \citenamefont {Hughes}, \citenamefont {Tan}, \citenamefont {Deng},
  \citenamefont {Sapra}, \citenamefont {England}, \citenamefont {Vuckovic},
  \citenamefont {Harris}, \citenamefont {Byer},\ and\ \citenamefont
  {Fan}}]{zhao2018design}%
  \BibitemOpen
  \bibfield  {author} {\bibinfo {author} {\bibfnamefont {Z.}~\bibnamefont
  {Zhao}}, \bibinfo {author} {\bibfnamefont {T.~W.}\ \bibnamefont {Hughes}},
  \bibinfo {author} {\bibfnamefont {S.}~\bibnamefont {Tan}}, \bibinfo {author}
  {\bibfnamefont {H.}~\bibnamefont {Deng}}, \bibinfo {author} {\bibfnamefont
  {N.}~\bibnamefont {Sapra}}, \bibinfo {author} {\bibfnamefont {R.~J.}\
  \bibnamefont {England}}, \bibinfo {author} {\bibfnamefont {J.}~\bibnamefont
  {Vuckovic}}, \bibinfo {author} {\bibfnamefont {J.~S.}\ \bibnamefont
  {Harris}}, \bibinfo {author} {\bibfnamefont {R.~L.}\ \bibnamefont {Byer}},
  and\ \bibinfo {author} {\bibfnamefont {S.}~\bibnamefont {Fan}},\ }\bibfield
  {title} {\bibinfo {title} {Design of a tapered slot waveguide dielectric
  laser accelerator for sub-relativistic electrons},\ }\href@noop {} {\bibfield
   {journal} {\bibinfo  {journal} {Optics Express}\ }\textbf {\bibinfo {volume}
  {26}},\ \bibinfo {pages} {22801} (\bibinfo {year} {2018})}\BibitemShut
  {NoStop}%
\bibitem [{\citenamefont {Sapra}\ \emph {et~al.}(2020)\citenamefont {Sapra},
  \citenamefont {Yang}, \citenamefont {Vercruysse}, \citenamefont {Leedle},
  \citenamefont {Black}, \citenamefont {England}, \citenamefont {Su},
  \citenamefont {Trivedi}, \citenamefont {Miao}, \citenamefont {Solgaard} \emph
  {et~al.}}]{sapra2019chip}%
  \BibitemOpen
  \bibfield  {author} {\bibinfo {author} {\bibfnamefont {N.~V.}\ \bibnamefont
  {Sapra}}, \bibinfo {author} {\bibfnamefont {K.~Y.}\ \bibnamefont {Yang}},
  \bibinfo {author} {\bibfnamefont {D.}~\bibnamefont {Vercruysse}}, \bibinfo
  {author} {\bibfnamefont {K.~J.}\ \bibnamefont {Leedle}}, \bibinfo {author}
  {\bibfnamefont {D.~S.}\ \bibnamefont {Black}}, \bibinfo {author}
  {\bibfnamefont {R.~J.}\ \bibnamefont {England}}, \bibinfo {author}
  {\bibfnamefont {L.}~\bibnamefont {Su}}, \bibinfo {author} {\bibfnamefont
  {R.}~\bibnamefont {Trivedi}}, \bibinfo {author} {\bibfnamefont
  {Y.}~\bibnamefont {Miao}}, \bibinfo {author} {\bibfnamefont {O.}~\bibnamefont
  {Solgaard}},  \emph {et~al.},\ }\bibfield  {title} {\bibinfo {title} {On-chip
  integrated laser-driven particle accelerator},\ }\href@noop {} {\bibfield
  {journal} {\bibinfo  {journal} {Science}\ }\textbf {\bibinfo {volume}
  {367}},\ \bibinfo {pages} {79} (\bibinfo {year} {2020})}\BibitemShut
  {NoStop}%
\bibitem [{\citenamefont {Niedermayer}\ \emph {et~al.}(2018)\citenamefont
  {Niedermayer}, \citenamefont {Egenolf}, \citenamefont {Boine-Frankenheim},\
  and\ \citenamefont {Hommelhoff}}]{niedermayer2018alternating}%
  \BibitemOpen
  \bibfield  {author} {\bibinfo {author} {\bibfnamefont {U.}~\bibnamefont
  {Niedermayer}}, \bibinfo {author} {\bibfnamefont {T.}~\bibnamefont
  {Egenolf}}, \bibinfo {author} {\bibfnamefont {O.}~\bibnamefont
  {Boine-Frankenheim}}, and\ \bibinfo {author} {\bibfnamefont {P.}~\bibnamefont
  {Hommelhoff}},\ }\bibfield  {title} {\bibinfo {title} {Alternating-phase
  focusing for dielectric-laser acceleration},\ }\href@noop {} {\bibfield
  {journal} {\bibinfo  {journal} {Physical Review Letters}\ }\textbf {\bibinfo
  {volume} {121}},\ \bibinfo {pages} {214801} (\bibinfo {year}
  {2018})}\BibitemShut {NoStop}%
\bibitem [{\citenamefont {Naranjo}\ \emph {et~al.}(2012)\citenamefont
  {Naranjo}, \citenamefont {Valloni}, \citenamefont {Putterman},\ and\
  \citenamefont {Rosenzweig}}]{naranjo2012stable}%
  \BibitemOpen
  \bibfield  {author} {\bibinfo {author} {\bibfnamefont {B.}~\bibnamefont
  {Naranjo}}, \bibinfo {author} {\bibfnamefont {A.}~\bibnamefont {Valloni}},
  \bibinfo {author} {\bibfnamefont {S.}~\bibnamefont {Putterman}}, and\
  \bibinfo {author} {\bibfnamefont {J.}~\bibnamefont {Rosenzweig}},\ }\bibfield
   {title} {\bibinfo {title} {Stable charged-particle acceleration and focusing
  in a laser accelerator using spatial harmonics},\ }\href@noop {} {\bibfield
  {journal} {\bibinfo  {journal} {Physical review letters}\ }\textbf {\bibinfo
  {volume} {109}},\ \bibinfo {pages} {164803} (\bibinfo {year}
  {2012})}\BibitemShut {NoStop}%
\bibitem [{\citenamefont {Wootton}\ \emph
  {et~al.}(2016{\natexlab{b}})\citenamefont {Wootton}, \citenamefont {McNeur},\
  and\ \citenamefont {Leedle}}]{wootton2016dielectric}%
  \BibitemOpen
  \bibfield  {author} {\bibinfo {author} {\bibfnamefont {K.}~\bibnamefont
  {Wootton}}, \bibinfo {author} {\bibfnamefont {J.}~\bibnamefont {McNeur}},
  and\ \bibinfo {author} {\bibfnamefont {K.}~\bibnamefont {Leedle}},\
  }\bibfield  {title} {\bibinfo {title} {Dielectric laser accelerators:
  designs, experiments, and applications},\ }\href@noop {} {\bibfield
  {journal} {\bibinfo  {journal} {Reviews of Accelerator Science and
  Technology}\ }\textbf {\bibinfo {volume} {9}},\ \bibinfo {pages} {105}
  (\bibinfo {year} {2016}{\natexlab{b}})}\BibitemShut {NoStop}%
\bibitem [{\citenamefont {Ody}\ \emph {et~al.}(2017)\citenamefont {Ody},
  \citenamefont {Musumeci}, \citenamefont {Maxson}, \citenamefont {Cesar},
  \citenamefont {England},\ and\ \citenamefont {Wootton}}]{ody2017flat}%
  \BibitemOpen
  \bibfield  {author} {\bibinfo {author} {\bibfnamefont {A.}~\bibnamefont
  {Ody}}, \bibinfo {author} {\bibfnamefont {P.}~\bibnamefont {Musumeci}},
  \bibinfo {author} {\bibfnamefont {J.}~\bibnamefont {Maxson}}, \bibinfo
  {author} {\bibfnamefont {D.}~\bibnamefont {Cesar}}, \bibinfo {author}
  {\bibfnamefont {R.}~\bibnamefont {England}}, and\ \bibinfo {author}
  {\bibfnamefont {K.}~\bibnamefont {Wootton}},\ }\bibfield  {title} {\bibinfo
  {title} {Flat electron beam sources for dla accelerators},\ }\href@noop {}
  {\bibfield  {journal} {\bibinfo  {journal} {Nuclear Instruments and Methods
  in Physics Research Section A: Accelerators, Spectrometers, Detectors and
  Associated Equipment}\ }\textbf {\bibinfo {volume} {865}},\ \bibinfo {pages}
  {75} (\bibinfo {year} {2017})}\BibitemShut {NoStop}%
\bibitem [{\citenamefont {Whittum}\ and\ \citenamefont
  {Tantawi}(2001)}]{whittum2001switched}%
  \BibitemOpen
  \bibfield  {author} {\bibinfo {author} {\bibfnamefont {D.~H.}\ \bibnamefont
  {Whittum}}and\ \bibinfo {author} {\bibfnamefont {S.~G.}\ \bibnamefont
  {Tantawi}},\ }\bibfield  {title} {\bibinfo {title} {Switched matrix
  accelerator},\ }\href@noop {} {\bibfield  {journal} {\bibinfo  {journal}
  {Review of Scientific Instruments}\ }\textbf {\bibinfo {volume} {72}},\
  \bibinfo {pages} {73} (\bibinfo {year} {2001})}\BibitemShut {NoStop}%
\bibitem [{\citenamefont {Whittum}\ and\ \citenamefont
  {Tantawi}(1998)}]{whittum1998active}%
  \BibitemOpen
  \bibfield  {author} {\bibinfo {author} {\bibfnamefont {D.~H.}\ \bibnamefont
  {Whittum}}and\ \bibinfo {author} {\bibfnamefont {S.~G.}\ \bibnamefont
  {Tantawi}},\ }\bibfield  {title} {\bibinfo {title} {Active millimeter wave
  accelerator with parallel beams},\ }in\ \href@noop {} {\emph {\bibinfo
  {booktitle} {SLAC-PUB-7845}}}\ (\bibinfo {year} {1998})\BibitemShut {NoStop}%
\bibitem [{\citenamefont {Zimmermann}\ \emph {et~al.}(1998)\citenamefont
  {Zimmermann}, \citenamefont {Hill},\ and\ \citenamefont
  {Whittum}}]{zimmermann1998new}%
  \BibitemOpen
  \bibfield  {author} {\bibinfo {author} {\bibfnamefont {F.}~\bibnamefont
  {Zimmermann}}, \bibinfo {author} {\bibfnamefont {M.}~\bibnamefont {Hill}},
  and\ \bibinfo {author} {\bibfnamefont {D.}~\bibnamefont {Whittum}},\
  }\bibfield  {title} {\bibinfo {title} {New concepts for a compact 5-{TeV}
  collider},\ }in\ \href@noop {} {\emph {\bibinfo {booktitle}
  {SLAC-PUB-7856}}}\ (\bibinfo {year} {1998})\ pp.\ \bibinfo {pages}
  {865--870}\BibitemShut {NoStop}%
\bibitem [{\citenamefont {Leedle}\ \emph {et~al.}(2018)\citenamefont {Leedle},
  \citenamefont {Black}, \citenamefont {Miao}, \citenamefont {Urbanek},
  \citenamefont {Ceballos}, \citenamefont {Deng}, \citenamefont {Harris},
  \citenamefont {Solgaard},\ and\ \citenamefont {Byer}}]{leedle2018phase}%
  \BibitemOpen
  \bibfield  {author} {\bibinfo {author} {\bibfnamefont {K.~J.}\ \bibnamefont
  {Leedle}}, \bibinfo {author} {\bibfnamefont {D.~S.}\ \bibnamefont {Black}},
  \bibinfo {author} {\bibfnamefont {Y.}~\bibnamefont {Miao}}, \bibinfo {author}
  {\bibfnamefont {K.~E.}\ \bibnamefont {Urbanek}}, \bibinfo {author}
  {\bibfnamefont {A.}~\bibnamefont {Ceballos}}, \bibinfo {author}
  {\bibfnamefont {H.}~\bibnamefont {Deng}}, \bibinfo {author} {\bibfnamefont
  {J.~S.}\ \bibnamefont {Harris}}, \bibinfo {author} {\bibfnamefont
  {O.}~\bibnamefont {Solgaard}}, and\ \bibinfo {author} {\bibfnamefont {R.~L.}\
  \bibnamefont {Byer}},\ }\bibfield  {title} {\bibinfo {title} {Phase-dependent
  laser acceleration of electrons with symmetrically driven silicon dual pillar
  gratings},\ }\href@noop {} {\bibfield  {journal} {\bibinfo  {journal} {Optics
  Letters}\ }\textbf {\bibinfo {volume} {43}},\ \bibinfo {pages} {2181}
  (\bibinfo {year} {2018})}\BibitemShut {NoStop}%
\bibitem [{\citenamefont {Black}\ \emph
  {et~al.}(2019{\natexlab{a}})\citenamefont {Black}, \citenamefont {Leedle},
  \citenamefont {Miao}, \citenamefont {Niedermayer}, \citenamefont {Byer},
  \citenamefont {Solgaard}, \citenamefont {Collaboration} \emph
  {et~al.}}]{black2019laser}%
  \BibitemOpen
  \bibfield  {author} {\bibinfo {author} {\bibfnamefont {D.~S.}\ \bibnamefont
  {Black}}, \bibinfo {author} {\bibfnamefont {K.~J.}\ \bibnamefont {Leedle}},
  \bibinfo {author} {\bibfnamefont {Y.}~\bibnamefont {Miao}}, \bibinfo {author}
  {\bibfnamefont {U.}~\bibnamefont {Niedermayer}}, \bibinfo {author}
  {\bibfnamefont {R.~L.}\ \bibnamefont {Byer}}, \bibinfo {author}
  {\bibfnamefont {O.}~\bibnamefont {Solgaard}}, \bibinfo {author}
  {\bibfnamefont {A.}~\bibnamefont {Collaboration}},  \emph {et~al.},\
  }\bibfield  {title} {\bibinfo {title} {Laser-driven electron lensing in
  silicon microstructures},\ }\href@noop {} {\bibfield  {journal} {\bibinfo
  {journal} {Physical Review Letters}\ }\textbf {\bibinfo {volume} {122}},\
  \bibinfo {pages} {104801} (\bibinfo {year} {2019}{\natexlab{a}})}\BibitemShut
  {NoStop}%
\bibitem [{\citenamefont {Joannopoulos}\ \emph {et~al.}(2008)\citenamefont
  {Joannopoulos}, \citenamefont {Johnson}, \citenamefont {Winn},\ and\
  \citenamefont {Meade}}]{joannopoulos2008molding}%
  \BibitemOpen
  \bibfield  {author} {\bibinfo {author} {\bibfnamefont {J.~D.}\ \bibnamefont
  {Joannopoulos}}, \bibinfo {author} {\bibfnamefont {S.~G.}\ \bibnamefont
  {Johnson}}, \bibinfo {author} {\bibfnamefont {J.~N.}\ \bibnamefont {Winn}},
  and\ \bibinfo {author} {\bibfnamefont {R.~D.}\ \bibnamefont {Meade}},\
  }\bibfield  {title} {\bibinfo {title} {Photonic crystals molding the flow of
  light},\ }\href@noop {} {\bibfield  {journal} {\bibinfo  {journal} {Princeton
  Univ. Press, Princeton, NJ}\ } (\bibinfo {year} {2008})}\BibitemShut
  {NoStop}%
\bibitem [{\citenamefont {Niedermayer}\ \emph {et~al.}(2017)\citenamefont
  {Niedermayer}, \citenamefont {Egenolf},\ and\ \citenamefont
  {Boine-Frankenheim}}]{niedermayer2017beam}%
  \BibitemOpen
  \bibfield  {author} {\bibinfo {author} {\bibfnamefont {U.}~\bibnamefont
  {Niedermayer}}, \bibinfo {author} {\bibfnamefont {T.}~\bibnamefont
  {Egenolf}}, and\ \bibinfo {author} {\bibfnamefont {O.}~\bibnamefont
  {Boine-Frankenheim}},\ }\bibfield  {title} {\bibinfo {title} {Beam dynamics
  analysis of dielectric laser acceleration using a fast 6d tracking scheme},\
  }\href@noop {} {\bibfield  {journal} {\bibinfo  {journal} {Physical Review
  Accelerators and Beams}\ }\textbf {\bibinfo {volume} {20}},\ \bibinfo {pages}
  {111302} (\bibinfo {year} {2017})}\BibitemShut {NoStop}%
\bibitem [{\citenamefont {Byer}(2013)}]{byer2013development}%
  \BibitemOpen
  \bibfield  {author} {\bibinfo {author} {\bibfnamefont {R.~L.}\ \bibnamefont
  {Byer}},\ }\href@noop {} {\emph {\bibinfo {title} {Development of
  high-gradient dielectric laser-driven particle accelerator structures}}},\
  \bibinfo {type} {Tech. Rep.}\ (\bibinfo  {institution} {Stanford Univ., CA
  (United States)},\ \bibinfo {year} {2013})\BibitemShut {NoStop}%
\bibitem [{\citenamefont {Wei}\ \emph {et~al.}(2017)\citenamefont {Wei},
  \citenamefont {Xia}, \citenamefont {Smith},\ and\ \citenamefont
  {Welsch}}]{wei2017dualbragg}%
  \BibitemOpen
  \bibfield  {author} {\bibinfo {author} {\bibfnamefont {Y.}~\bibnamefont
  {Wei}}, \bibinfo {author} {\bibfnamefont {G.}~\bibnamefont {Xia}}, \bibinfo
  {author} {\bibfnamefont {J.}~\bibnamefont {Smith}}, and\ \bibinfo {author}
  {\bibfnamefont {C.}~\bibnamefont {Welsch}},\ }\bibfield  {title} {\bibinfo
  {title} {Dual-gratings with a bragg reflector for dielectric laser-driven
  accelerators},\ }\href@noop {} {\bibfield  {journal} {\bibinfo  {journal}
  {Physics of Plasmas}\ }\textbf {\bibinfo {volume} {24}},\ \bibinfo {pages}
  {073115} (\bibinfo {year} {2017})}\BibitemShut {NoStop}%
\bibitem [{\citenamefont {Miao}\ \emph {et~al.}(2020)\citenamefont {Miao},
  \citenamefont {Black}, \citenamefont {Leedle}, \citenamefont {Zhao},
  \citenamefont {Deng}, \citenamefont {Ceballos}, \citenamefont {Byer},
  \citenamefont {Harris},\ and\ \citenamefont {Solgaard}}]{Miao2019surface}%
  \BibitemOpen
  \bibfield  {author} {\bibinfo {author} {\bibfnamefont {Y.}~\bibnamefont
  {Miao}}, \bibinfo {author} {\bibfnamefont {D.~S.}\ \bibnamefont {Black}},
  \bibinfo {author} {\bibfnamefont {K.~J.}\ \bibnamefont {Leedle}}, \bibinfo
  {author} {\bibfnamefont {Z.}~\bibnamefont {Zhao}}, \bibinfo {author}
  {\bibfnamefont {H.}~\bibnamefont {Deng}}, \bibinfo {author} {\bibfnamefont
  {A.}~\bibnamefont {Ceballos}}, \bibinfo {author} {\bibfnamefont {R.~L.}\
  \bibnamefont {Byer}}, \bibinfo {author} {\bibfnamefont {J.~S.}\ \bibnamefont
  {Harris}}, and\ \bibinfo {author} {\bibfnamefont {O.}~\bibnamefont
  {Solgaard}},\ }\bibfield  {title} {\bibinfo {title} {Surface treatments of
  dielectric laser accelerators for increased laser-induced damage threshold},\
  }\href@noop {} {\bibfield  {journal} {\bibinfo  {journal} {Optics Letters}\
  }\textbf {\bibinfo {volume} {45}},\ \bibinfo {pages} {391} (\bibinfo {year}
  {2020})}\BibitemShut {NoStop}%
\bibitem [{\citenamefont {Shin}\ and\ \citenamefont
  {Fan}(2012)}]{shin2012choice}%
  \BibitemOpen
  \bibfield  {author} {\bibinfo {author} {\bibfnamefont {W.}~\bibnamefont
  {Shin}}, and\ \bibinfo {author} {\bibfnamefont {S.}~\bibnamefont {Fan}},\
  }\bibfield  {title} {\bibinfo {title} {Choice of the perfectly matched layer
  boundary condition for frequency-domain maxwell’s equations solvers},\
  }\href@noop {} {\bibfield  {journal} {\bibinfo  {journal} {Journal of
  Computational Physics}\ }\textbf {\bibinfo {volume} {231}},\ \bibinfo {pages}
  {3406} (\bibinfo {year} {2012})}\BibitemShut {NoStop}%
\bibitem [{\citenamefont {Lee}(2017)}]{lee2017ultrafast}%
  \BibitemOpen
  \bibfield  {author} {\bibinfo {author} {\bibfnamefont {Y.~J.}\ \bibnamefont
  {Lee}},\ }\emph {\bibinfo {title} {Ultrafast laser-induced damage threshold
  of the optical materials in near-infrared region}},\ \href@noop {} {Ph.D.
  thesis},\ \bibinfo  {school} {Purdue University} (\bibinfo {year}
  {2017})\BibitemShut {NoStop}%
\bibitem [{\citenamefont {Sch{\"a}chter}\ \emph {et~al.}(2002)\citenamefont
  {Sch{\"a}chter}, \citenamefont {Byer},\ and\ \citenamefont
  {Siemann}}]{schachter2002optical}%
  \BibitemOpen
  \bibfield  {author} {\bibinfo {author} {\bibfnamefont {L.}~\bibnamefont
  {Sch{\"a}chter}}, \bibinfo {author} {\bibfnamefont {R.~L.}\ \bibnamefont
  {Byer}}, and\ \bibinfo {author} {\bibfnamefont {R.~H.}\ \bibnamefont
  {Siemann}},\ }\bibfield  {title} {\bibinfo {title} {Optical accelerator:
  Scaling laws and figures of merit},\ }in\ \href@noop {} {\emph {\bibinfo
  {booktitle} {AIP Conference Proceedings}}},\ Vol.\ \bibinfo {volume} {647}\
  (\bibinfo {organization} {AIP},\ \bibinfo {year} {2002})\ pp.\ \bibinfo
  {pages} {310--323}\BibitemShut {NoStop}%
\bibitem [{\citenamefont {Siemann}(2004)}]{siemann2004energy}%
  \BibitemOpen
  \bibfield  {author} {\bibinfo {author} {\bibfnamefont {R.}~\bibnamefont
  {Siemann}},\ }\bibfield  {title} {\bibinfo {title} {Energy efficiency of
  laser driven, structure based accelerators},\ }\href@noop {} {\bibfield
  {journal} {\bibinfo  {journal} {Physical Review Special Topics-Accelerators
  and Beams}\ }\textbf {\bibinfo {volume} {7}},\ \bibinfo {pages} {061303}
  (\bibinfo {year} {2004})}\BibitemShut {NoStop}%
\bibitem [{\citenamefont {McNeur}\ \emph {et~al.}(2018)\citenamefont {McNeur},
  \citenamefont {Koz{\'a}k}, \citenamefont {Sch{\"o}nenberger}, \citenamefont
  {Leedle}, \citenamefont {Deng}, \citenamefont {Ceballos}, \citenamefont
  {Hoogland}, \citenamefont {Ruehl}, \citenamefont {Hartl}, \citenamefont
  {Holzwarth} \emph {et~al.}}]{mcneur2018elements}%
  \BibitemOpen
  \bibfield  {author} {\bibinfo {author} {\bibfnamefont {J.}~\bibnamefont
  {McNeur}}, \bibinfo {author} {\bibfnamefont {M.}~\bibnamefont {Koz{\'a}k}},
  \bibinfo {author} {\bibfnamefont {N.}~\bibnamefont {Sch{\"o}nenberger}},
  \bibinfo {author} {\bibfnamefont {K.~J.}\ \bibnamefont {Leedle}}, \bibinfo
  {author} {\bibfnamefont {H.}~\bibnamefont {Deng}}, \bibinfo {author}
  {\bibfnamefont {A.}~\bibnamefont {Ceballos}}, \bibinfo {author}
  {\bibfnamefont {H.}~\bibnamefont {Hoogland}}, \bibinfo {author}
  {\bibfnamefont {A.}~\bibnamefont {Ruehl}}, \bibinfo {author} {\bibfnamefont
  {I.}~\bibnamefont {Hartl}}, \bibinfo {author} {\bibfnamefont
  {R.}~\bibnamefont {Holzwarth}},  \emph {et~al.},\ }\bibfield  {title}
  {\bibinfo {title} {Elements of a dielectric laser accelerator},\ }\href@noop
  {} {\bibfield  {journal} {\bibinfo  {journal} {Optica}\ }\textbf {\bibinfo
  {volume} {5}},\ \bibinfo {pages} {687} (\bibinfo {year} {2018})}\BibitemShut
  {NoStop}%
\bibitem [{\citenamefont {Tiberio}\ \emph {et~al.}(1998)\citenamefont
  {Tiberio}, \citenamefont {Carr}, \citenamefont {Rooks}, \citenamefont
  {Mihailov}, \citenamefont {Bilodeau}, \citenamefont {Albert}, \citenamefont
  {Stryckman}, \citenamefont {Johnson}, \citenamefont {Hill}, \citenamefont
  {McClelland} \emph {et~al.}}]{tiberio1998fabrication}%
  \BibitemOpen
  \bibfield  {author} {\bibinfo {author} {\bibfnamefont {R.}~\bibnamefont
  {Tiberio}}, \bibinfo {author} {\bibfnamefont {D.}~\bibnamefont {Carr}},
  \bibinfo {author} {\bibfnamefont {M.}~\bibnamefont {Rooks}}, \bibinfo
  {author} {\bibfnamefont {S.}~\bibnamefont {Mihailov}}, \bibinfo {author}
  {\bibfnamefont {F.}~\bibnamefont {Bilodeau}}, \bibinfo {author}
  {\bibfnamefont {J.}~\bibnamefont {Albert}}, \bibinfo {author} {\bibfnamefont
  {D.}~\bibnamefont {Stryckman}}, \bibinfo {author} {\bibfnamefont
  {D.}~\bibnamefont {Johnson}}, \bibinfo {author} {\bibfnamefont
  {K.}~\bibnamefont {Hill}}, \bibinfo {author} {\bibfnamefont {A.}~\bibnamefont
  {McClelland}},  \emph {et~al.},\ }\bibfield  {title} {\bibinfo {title}
  {Fabrication of electron beam generated, chirped, phase mask
  (1070.11--1070.66 nm) for fiber bragg grating dispersion compensator},\
  }\href@noop {} {\bibfield  {journal} {\bibinfo  {journal} {Journal of Vacuum
  Science \& Technology B: Microelectronics and Nanometer Structures
  Processing, Measurement, and Phenomena}\ }\textbf {\bibinfo {volume} {16}},\
  \bibinfo {pages} {3237} (\bibinfo {year} {1998})}\BibitemShut {NoStop}%
\bibitem [{\citenamefont {Black}\ \emph
  {et~al.}(2019{\natexlab{b}})\citenamefont {Black}, \citenamefont
  {Niedermayer}, \citenamefont {Miao}, \citenamefont {Zhao}, \citenamefont
  {Solgaard}, \citenamefont {Byer},\ and\ \citenamefont
  {Leedle}}]{black2019generation}%
  \BibitemOpen
  \bibfield  {author} {\bibinfo {author} {\bibfnamefont {D.~S.}\ \bibnamefont
  {Black}}, \bibinfo {author} {\bibfnamefont {U.}~\bibnamefont {Niedermayer}},
  \bibinfo {author} {\bibfnamefont {Y.}~\bibnamefont {Miao}}, \bibinfo {author}
  {\bibfnamefont {Z.}~\bibnamefont {Zhao}}, \bibinfo {author} {\bibfnamefont
  {O.}~\bibnamefont {Solgaard}}, \bibinfo {author} {\bibfnamefont {R.~L.}\
  \bibnamefont {Byer}}, and\ \bibinfo {author} {\bibfnamefont {K.~J.}\
  \bibnamefont {Leedle}},\ }\bibfield  {title} {\bibinfo {title} {Net
  acceleration and direct measurement of attosecond electron pulses in a
  silicon dielectric laser accelerator},\ }\href@noop {} {\bibfield  {journal}
  {\bibinfo  {journal} {Physical Review Letters}\ }\textbf {\bibinfo {volume}
  {123}},\ \bibinfo {pages} {264802} (\bibinfo {year}
  {2019}{\natexlab{b}})}\BibitemShut {NoStop}%
\bibitem [{\citenamefont {Grote}\ \emph {et~al.}(2003)\citenamefont {Grote},
  \citenamefont {Henestroza},\ and\ \citenamefont {Kwan}}]{grote2003design}%
  \BibitemOpen
  \bibfield  {author} {\bibinfo {author} {\bibfnamefont {D.~P.}\ \bibnamefont
  {Grote}}, \bibinfo {author} {\bibfnamefont {E.}~\bibnamefont {Henestroza}},
  and\ \bibinfo {author} {\bibfnamefont {J.~W.}\ \bibnamefont {Kwan}},\
  }\bibfield  {title} {\bibinfo {title} {Design and simulation of a
  multibeamlet injector for a high current accelerator},\ }\href@noop {}
  {\bibfield  {journal} {\bibinfo  {journal} {Physical Review Special
  Topics-Accelerators and Beams}\ }\textbf {\bibinfo {volume} {6}},\ \bibinfo
  {pages} {014202} (\bibinfo {year} {2003})}\BibitemShut {NoStop}%
\bibitem [{\citenamefont {Yang}\ \emph {et~al.}(2018)\citenamefont {Yang},
  \citenamefont {Massuda}, \citenamefont {Roques-Carmes}, \citenamefont {Kooi},
  \citenamefont {Christensen}, \citenamefont {Johnson}, \citenamefont
  {Joannopoulos}, \citenamefont {Miller}, \citenamefont {Kaminer},\ and\
  \citenamefont {Solja{\v{c}}i{\'c}}}]{yang2018maximal}%
  \BibitemOpen
  \bibfield  {author} {\bibinfo {author} {\bibfnamefont {Y.}~\bibnamefont
  {Yang}}, \bibinfo {author} {\bibfnamefont {A.}~\bibnamefont {Massuda}},
  \bibinfo {author} {\bibfnamefont {C.}~\bibnamefont {Roques-Carmes}}, \bibinfo
  {author} {\bibfnamefont {S.~E.}\ \bibnamefont {Kooi}}, \bibinfo {author}
  {\bibfnamefont {T.}~\bibnamefont {Christensen}}, \bibinfo {author}
  {\bibfnamefont {S.~G.}\ \bibnamefont {Johnson}}, \bibinfo {author}
  {\bibfnamefont {J.~D.}\ \bibnamefont {Joannopoulos}}, \bibinfo {author}
  {\bibfnamefont {O.~D.}\ \bibnamefont {Miller}}, \bibinfo {author}
  {\bibfnamefont {I.}~\bibnamefont {Kaminer}}, and\ \bibinfo {author}
  {\bibfnamefont {M.}~\bibnamefont {Solja{\v{c}}i{\'c}}},\ }\bibfield  {title}
  {\bibinfo {title} {Maximal spontaneous photon emission and energy loss from
  free electrons},\ }\href@noop {} {\bibfield  {journal} {\bibinfo  {journal}
  {Nature Physics}\ }\textbf {\bibinfo {volume} {14}},\ \bibinfo {pages} {894}
  (\bibinfo {year} {2018})}\BibitemShut {NoStop}%
\bibitem [{\citenamefont {Zhao}\ \emph {et~al.}(2019)\citenamefont {Zhao},
  \citenamefont {Williamson}, \citenamefont {Zhao}, \citenamefont {Boutami},\
  and\ \citenamefont {Fan}}]{zhao2019penetration}%
  \BibitemOpen
  \bibfield  {author} {\bibinfo {author} {\bibfnamefont {N.~Z.}\ \bibnamefont
  {Zhao}}, \bibinfo {author} {\bibfnamefont {I.~A.}\ \bibnamefont
  {Williamson}}, \bibinfo {author} {\bibfnamefont {Z.}~\bibnamefont {Zhao}},
  \bibinfo {author} {\bibfnamefont {S.}~\bibnamefont {Boutami}}, and\ \bibinfo
  {author} {\bibfnamefont {S.}~\bibnamefont {Fan}},\ }\bibfield  {title}
  {\bibinfo {title} {Penetration depth reduction with plasmonic metafilms},\
  }\href@noop {} {\bibfield  {journal} {\bibinfo  {journal} {ACS Photonics}\
  }\textbf {\bibinfo {volume} {6}},\ \bibinfo {pages} {2049} (\bibinfo {year}
  {2019})}\BibitemShut {NoStop}%
\bibitem [{\citenamefont {Sousa}(2009)}]{sousa2009dose}%
  \BibitemOpen
  \bibfield  {author} {\bibinfo {author} {\bibfnamefont {R.}~\bibnamefont
  {Sousa}},\ }\bibfield  {title} {\bibinfo {title} {Dose rate influence on deep
  dose deposition using a 6 {MV} x-ray beam from a linear accelerator},\
  }\href@noop {} {\bibfield  {journal} {\bibinfo  {journal} {Brazilian Journal
  of Physics}\ }\textbf {\bibinfo {volume} {39}},\ \bibinfo {pages} {292}
  (\bibinfo {year} {2009})}\BibitemShut {NoStop}%
\bibitem [{\citenamefont {Tafel}\ \emph {et~al.}(2019)\citenamefont {Tafel},
  \citenamefont {Ristein},\ and\ \citenamefont
  {Hommelhoff}}]{tafel2019femtosecond}%
  \BibitemOpen
  \bibfield  {author} {\bibinfo {author} {\bibfnamefont {A.}~\bibnamefont
  {Tafel}}, \bibinfo {author} {\bibfnamefont {J.}~\bibnamefont {Ristein}}, and\
  \bibinfo {author} {\bibfnamefont {P.}~\bibnamefont {Hommelhoff}},\ }\bibfield
   {title} {\bibinfo {title} {Femtosecond laser-induced electron emission from
  nanodiamond-coated tungsten needle tips},\ }\href@noop {} {\bibfield
  {journal} {\bibinfo  {journal} {arXiv preprint arXiv:1903.05560}\ } (\bibinfo
  {year} {2019})}\BibitemShut {NoStop}%
\bibitem [{\citenamefont {Luo}\ \emph {et~al.}(2003)\citenamefont {Luo},
  \citenamefont {Ibanescu}, \citenamefont {Johnson},\ and\ \citenamefont
  {Joannopoulos}}]{luo2003cerenkov}%
  \BibitemOpen
  \bibfield  {author} {\bibinfo {author} {\bibfnamefont {C.}~\bibnamefont
  {Luo}}, \bibinfo {author} {\bibfnamefont {M.}~\bibnamefont {Ibanescu}},
  \bibinfo {author} {\bibfnamefont {S.~G.}\ \bibnamefont {Johnson}}, and\
  \bibinfo {author} {\bibfnamefont {J.}~\bibnamefont {Joannopoulos}},\
  }\bibfield  {title} {\bibinfo {title} {Cerenkov radiation in photonic
  crystals},\ }\href@noop {} {\bibfield  {journal} {\bibinfo  {journal}
  {science}\ }\textbf {\bibinfo {volume} {299}},\ \bibinfo {pages} {368}
  (\bibinfo {year} {2003})}\BibitemShut {NoStop}%
\bibitem [{\citenamefont {Ceballos}()}]{ceballos2014silicon}%
  \BibitemOpen
  \bibfield  {author} {\bibinfo {author} {\bibfnamefont {A.}~\bibnamefont
  {Ceballos}},\ }\bibfield  {title} {\bibinfo {title} {{Silicon Microstructures
  for Electron Acceleration}},\ }in\ \href@noop {} {\emph {\bibinfo {booktitle}
  {{{Proceedings of the Advanced Accelerator Concepts Workshop, San Jose, CA,
  16 July 2014}}}}}\BibitemShut {NoStop}%
\bibitem [{\citenamefont {Sears}\ \emph {et~al.}(2008)\citenamefont {Sears},
  \citenamefont {Colby}, \citenamefont {Ischebeck}, \citenamefont {McGuinness},
  \citenamefont {Nelson}, \citenamefont {Noble}, \citenamefont {Siemann},
  \citenamefont {Spencer}, \citenamefont {Walz}, \citenamefont {Plettner} \emph
  {et~al.}}]{sears2008production}%
  \BibitemOpen
  \bibfield  {author} {\bibinfo {author} {\bibfnamefont {C.~M.}\ \bibnamefont
  {Sears}}, \bibinfo {author} {\bibfnamefont {E.}~\bibnamefont {Colby}},
  \bibinfo {author} {\bibfnamefont {R.}~\bibnamefont {Ischebeck}}, \bibinfo
  {author} {\bibfnamefont {C.}~\bibnamefont {McGuinness}}, \bibinfo {author}
  {\bibfnamefont {J.}~\bibnamefont {Nelson}}, \bibinfo {author} {\bibfnamefont
  {R.}~\bibnamefont {Noble}}, \bibinfo {author} {\bibfnamefont {R.~H.}\
  \bibnamefont {Siemann}}, \bibinfo {author} {\bibfnamefont {J.}~\bibnamefont
  {Spencer}}, \bibinfo {author} {\bibfnamefont {D.}~\bibnamefont {Walz}},
  \bibinfo {author} {\bibfnamefont {T.}~\bibnamefont {Plettner}},  \emph
  {et~al.},\ }\bibfield  {title} {\bibinfo {title} {Production and
  characterization of attosecond electron bunch trains},\ }\href@noop {}
  {\bibfield  {journal} {\bibinfo  {journal} {Physical Review Special
  Topics-Accelerators and Beams}\ }\textbf {\bibinfo {volume} {11}},\ \bibinfo
  {pages} {061301} (\bibinfo {year} {2008})}\BibitemShut {NoStop}%
\bibitem [{\citenamefont {Sch{\"o}nenberger}\ \emph {et~al.}(2019)\citenamefont
  {Sch{\"o}nenberger}, \citenamefont {Mittelbach}, \citenamefont {Yousefi},
  \citenamefont {McNeur}, \citenamefont {Niedermayer},\ and\ \citenamefont
  {Hommelhoff}}]{schonenberger2019generation}%
  \BibitemOpen
  \bibfield  {author} {\bibinfo {author} {\bibfnamefont {N.}~\bibnamefont
  {Sch{\"o}nenberger}}, \bibinfo {author} {\bibfnamefont {A.}~\bibnamefont
  {Mittelbach}}, \bibinfo {author} {\bibfnamefont {P.}~\bibnamefont {Yousefi}},
  \bibinfo {author} {\bibfnamefont {J.}~\bibnamefont {McNeur}}, \bibinfo
  {author} {\bibfnamefont {U.}~\bibnamefont {Niedermayer}}, and\ \bibinfo
  {author} {\bibfnamefont {P.}~\bibnamefont {Hommelhoff}},\ }\bibfield  {title}
  {\bibinfo {title} {Generation and characterization of attosecond microbunched
  electron pulse trains via dielectric laser acceleration},\ }\href@noop {}
  {\bibfield  {journal} {\bibinfo  {journal} {Physical Review Letters}\
  }\textbf {\bibinfo {volume} {123}},\ \bibinfo {pages} {264803} (\bibinfo
  {year} {2019})}\BibitemShut {NoStop}%
\end{thebibliography}%

\end{document}